\documentclass[reprint,prr,longbibliography,superscriptaddress,nofootinbib]{revtex4-2}
\usepackage{lipsum}
\usepackage{fancyhdr}
\usepackage{graphicx}
\usepackage[caption=false]{subfig}
\usepackage{braket}
\usepackage[dvipsnames]{xcolor}
\usepackage{mathtools}
\usepackage{amsmath}
\usepackage{amssymb}
\usepackage{hyperref}
\usepackage{esvect}
\usepackage{float}
\usepackage{notes2bib}
\graphicspath{{./Figures/}}

\usepackage{amsmath}
\usepackage{multirow}

\usepackage{amsfonts}
\usepackage{bm}
\usepackage{bbold}

\hypersetup{
    colorlinks=true,
    linkcolor=blue,
    filecolor=magenta,      
    urlcolor=magenta,
    citecolor={blue},
    }

\allowdisplaybreaks

\usepackage{letltxmacro}
\LetLtxMacro{\ORIGselectlanguage}{\selectlanguage}
\makeatletter
\DeclareRobustCommand{\selectlanguage}[1]{%
  \@ifundefined{alias@\string#1}
    {\ORIGselectlanguage{#1}}
    {\begingroup\edef\x{\endgroup
       \noexpand\ORIGselectlanguage{\@nameuse{alias@#1}}}\x}%
}
\newcommand{\definelanguagealias}[2]{%
  \@namedef{alias@#1}{#2}%
}

\makeatother

\definelanguagealias{en}{english}
\definelanguagealias{EN}{english}
\definelanguagealias{eng}{english}
\definelanguagealias{de}{ngerman}

\begin{document}

\title{Classical Purcell factors 
 and spontaneous emission decay rates in a linear gain medium}

\author{Juanjuan Ren}
\email{jr180@queensu.ca}
\affiliation{Department of Physics, Engineering Physics, and Astronomy, Queen's
University, Kingston, Ontario K7L 3N6, Canada}
\author{Sebastian Franke}
\affiliation{Department of Physics, Engineering Physics, and Astronomy, Queen's
University, Kingston, Ontario K7L 3N6, Canada}
\affiliation{Technische Universit\"at Berlin, Institut f\"ur Theoretische Physik, Nichtlineare Optik und Quantenelektronik, Hardenbergstra{\ss}e 36, 10623 Berlin, Germany}
\author{Becca VanDrunen}
\affiliation{Department of Physics, Engineering Physics, and Astronomy, Queen's
University, Kingston, Ontario K7L 3N6, Canada}
\author{Stephen Hughes}
\affiliation{Department of Physics, Engineering Physics, and Astronomy, Queen's
University, Kingston, Ontario K7L 3N6, Canada}

\begin{abstract}
Recently the photonic golden rule, which predicts that the spontaneous emission rate of an atom depends on the projected local density of states (LDOS), was shown to fail in an optical medium with a linear gain amplifier. We present a classical light-matter theory to fix this widely used spontaneous emission rate, fully recovering the quantum mechanical rate reported in Franke et al., Phys. Rev. Lett. 127, 013602 (2021). The corrected classical Purcell factor, for media containing linear amplifiers, is obtained in two different forms, both of which can easily be calculated in any standard classical Maxwell solver. We also derive explicit analytical results in terms of quasinormal modes, which are useful for studying practical cavity structures in an efficient way, {including the presence of local field effects for finite-size dipole emitters embedded inside lossy or gain materials (using a real cavity model). Finally, we derive a full classical correspondence from the viewpoint of quantized quasinormal modes in the bad cavity limit. Example numerical calculations are shown for coupled loss-gain microdisk resonators, showing excellent agreement between few mode expansions and full numerical dipole simulations.}

\end{abstract}
\maketitle

\section{Introduction}

Spontaneous emission (SE) is one of the most striking examples of quantum electrodynamics (QED)~\cite{dirac1927quantum,Weisskopf1930,RevModPhys.4.87}, where vacuum fluctuations cause the emission of a photon in the absence of any coherence, from an excited state population~\cite{Milonni1976,PhysRevLett.77.2444,dung2000spontaneous,GirishBook1}.
On the other hand, a classical interpretation of 
SE is also possible, whereby a
classical dipole is excited, and the subsequent SE decay can be interpreted through ``radiation reaction''~\cite{Barnes2020}. 
In a fully quantum description,
both viewpoints are valid, depending on the chosen ordering of the operators, including mixed operator ordering~\cite{Milonni1976}.
This is extremely convenient for modelling SE in various nanophotonic cavity environments~\cite{PhysRevLett.58.2059,burstein_controlling_1995,lodahl_controlling_2004}, and is well exploited in a number of research fields, since classical field solvers can be used to obtain such rates.
Understanding and controlling SE is also of fundamental interest to many
areas in nanophotonics~\cite{Novotny2007Principles}, including the study
of nano-lasers~\cite{chow_emission_2014,Deng2021,lippi_amplified_2021}, 
 active fibers~\cite{PhysRevA.64.033812}, exceptional points~\cite{lin_enhanced_2016,pick_general_2017,zdemir2019,miri_exceptional_2019,khanbekyan_decay_2020,PhysRevA.100.062131,ferrier_unveiling_2022}, and coupled loss-gain systems~\cite{peng_paritytime-symmetric_2014,chang_paritytime_2014,liu_efficient_2011,ren_quasinormal_2021}.

For a point-dipole emitter
or two-level system (TLS) at
position ${\bf r}_{\rm 0}$, with
dipole moment, ${\bf d}$ (assumed real),
the SE rate in a {\it lossy medium} (including also lossless dielectrics) takes on the following form:
\begin{align}
\Gamma^{\rm SE}(\mathbf{r}_{0},\omega) &= \frac{\pi \omega \vert\mathbf{d}\vert^2}{3\hbar\epsilon_0} \rho^{\rm LDOS}(\mathbf{r}_{0},\omega) 
\equiv \Gamma^{\rm LDOS}(\mathbf{r}_{0},\omega)
\nonumber \\
&=\frac{2}{\hbar\epsilon_0}\mathbf{d}\cdot {\rm Im}\left[{\mathbf{G}}(\mathbf{r}_{0},\mathbf{r}_{0},\omega)\right]\cdot \mathbf{d},\label{eq: Gamma_LDOS}
\end{align}
where $\rho^{\rm LDOS}$ is the (projected) local density of
states (LDOS), and $\mathbf{G}$ is the classical Green function
of the medium, defined from
\begin{equation}
\boldsymbol{\nabla}\times\boldsymbol{\nabla}\times\mathbf{G}(\mathbf{r},\mathbf{r}',\omega) -\frac{\omega^2}{c^2}\epsilon(\mathbf{r},\omega)\mathbf{G}(\mathbf{r},\mathbf{r}',\omega)=\frac{\omega^2}{c^2}\mathbb{1}\delta(\mathbf{r}-\mathbf{r}'),
\label{eq: GFHelmholtz2}
\end{equation}
and $\epsilon({\bf r},\omega)$
is the dielectric function of the medium
which is in general complex, and ${\mathbb{1}}$ is the unit tensor.
This picture of SE decay can be derived using standard perturbation theories, and is valid when the emitter-medium coupling is weak, i.e., when the field and matter states are not entangled.

For single mode cavities, and for dipoles
aligned with the cavity mode polarization, at a field maximum, the standard Purcell formula can be written as
\begin{equation}
    F_{\rm P} \equiv \frac{3}{4 \pi^2}\left(\frac{\lambda}{n_{\rm B}}\right)^3\frac{Q}{V_{\rm eff}},
    \label{eq:original_pf}
\end{equation}
where $\lambda$ is the wavelength,
$n_{\rm B}$ is the background refractive index,
$Q$ is the quality factor, and 
 $V_{\rm eff}$ is the effective mode volume~\cite{kristensen_modes_2014}.
A more general Purcell factor
can be defined through~\cite{kristensen_modes_2014,ge_quasinormal_2014}
\begin{equation}
F_{\rm P}^{\rm LDOS}({\bf r}_0,\omega)
= 1+ \frac{\Gamma^{\rm SE}({\bf r}_0,\omega)}{\Gamma_0(\omega)},
\end{equation}
where 
\begin{equation}
\Gamma_0(\omega)=\frac{2}{\hbar\epsilon_0}\mathbf{d}\cdot {\rm Im}\left[{\mathbf{G}_{\rm hom}}(\mathbf{r}_{0},\mathbf{r}_{0},\omega)\right]\cdot \mathbf{d},
\end{equation}
is the homogeneous medium SE rate, 
obtained using the background Green function: ${\rm Im}[{\bf G}_{\rm hom}(\mathbf{r}_{0},\mathbf{r}_{0},\omega)] = {\mathbb{1}} n_{\rm B}\omega^3/(6 \pi c^3)$, with $\Gamma^{\rm SE}({\bf r}_0,\omega)$ defined from Eq.~\eqref{eq: Gamma_LDOS}.
For a 2D TM system,
as we will consider below for our numerical example, ${\rm Im}[{\bf G}_{\rm hom}(\mathbf{r}_{0},\mathbf{r}_{0},\omega)] = {\mathbb{1}} \omega^2/(4 c^2)$.
For dipoles that are located outside the
cavity structure of interest,  then we also include
a factor of 1~\cite{ge_quasinormal_2014};
otherwise the factor of 1 can be dropped.

To appreciate how vacuum fluctuations connect to the LDOS { in dielectric media, we briefly sketch out a quantum mechanical derivation of the SE rate. We start by using a Green function approach~\cite{Dung} for quantizing the electric field in a purely lossy medium, yielding $\hat{\bf E}({\bf r})=\int_0^\infty {\rm d}\omega \hat{\bf E}({\bf r},\omega)+{\rm H.a.}$, with
\begin{equation}
\hat{\bf E}({\bf r},\omega)=\frac{i}{\omega\epsilon_0}\int d{\bf r}' {\bf G}({\bf r},{\bf r}',\omega)
\cdot \hat {\bf j}({\bf r}',\omega), \label{eq: Eop_lossy}
\end{equation}
where $\hat {\bf j}({\bf r},\omega)\propto \sqrt{\epsilon_{\rm Im}(\mathbf{r},\omega)}\hat {\bf b}({\bf r},\omega)$ is a current noise operator counteracting the non-radiative and radiative dissipation, which are encoded in the imaginary part $\epsilon_{\rm Im}$ of the complex permittivity $\epsilon({\bf r},\omega)=\epsilon_{\rm Re}(\mathbf{r},\omega)+i\epsilon_{\rm Im}(\mathbf{r},\omega)$ (with $\epsilon_{\rm Im}(\mathbf{r},\omega)>0)$ and an integration over all space from Eq.~\eqref{eq: Eop_lossy} [for more details on the derivation of the radiative dissipation, cf. Ref.~\onlinecite{franke_fluctuation-dissipation_2020}]. Moreover, $\hat {\bf b}^{(\dagger)}({\bf r},\omega)$ are the bosonic operators of the combined medium-photon system and act as annihilation (creation) operator on the Fock states $|{\bf n}\rangle$ (containing all spatial and frequency dependencies), respectively. The point-dipole emitter is introduced into the quantum model as a simple two-level system and is coupled to the electric field via dipole-field interaction Hamiltonian $\hat{H}_I=\hat{\mathbf{d}}\cdot\hat{\mathbf{E}}(\mathbf{r}_0)$. Here, $\hat{\mathbf{d}}$ is the dipole operator $\hat{{\bf d}}=\mathbf{d}\hat{\sigma}^+ + \mathbf{d}\hat{\sigma}^-$ and $\hat{\sigma}^{-(+)}$ is the lowering (raising) operator, acting on the ground state $|g\rangle$ and the excited state $|e\rangle$. Next, we will} consider
the SE rate as defined through vacuum fluctuations~\cite{Milonni1976}
of the quantum field
\begin{equation}
    \Gamma^{\rm VF}({\bf r}_{\rm 0},\omega)
    =\frac{2\pi}{\hbar^2}\mathbf{d}\cdot\langle 0|[\hat{\mathbf{E}}(\mathbf{r}_{\rm 0},\omega_{\rm }),\hat{\mathbf{E}}^\dagger(\mathbf{r}_{\rm 0},\omega_{\rm }) ]|0\rangle\cdot\mathbf{d},\label{eq: SEClass2}
\end{equation}
 where $|0\rangle$ is the {photon-medium} vacuum state{, which fulfils $\hat {\bf b}({\bf r},\omega)|0\rangle = 0$. 
 Inserting the quantum field from Eq.~\eqref{eq: Eop_lossy}, one immediately recognizes that} this rate explicitly depends on an integration over all space. However, exploiting the
 Green function identity~\cite{Dung,franke_fluctuation-dissipation_2020} 
 \begin{equation}
 \int_{\mathbb{R}^3}{\rm d}\mathbf{s}\epsilon_{\rm Im}(\mathbf{s},\omega)\mathbf{G}(\mathbf{r},\mathbf{s},\omega)\cdot\mathbf{G}^*(\mathbf{s},\mathbf{r}',\omega)={\rm Im}[\mathbf{G}(\mathbf{r},\mathbf{r}',\omega)],
 \end{equation}
 then one can show that
 $\Gamma^{\rm VF} = \Gamma^{\rm LDOS} = \Gamma^{\rm SE}$, and both approaches yield equivalent results for the SE decay rate, which depends on the projected LDOS. Consequently, one can also model the enhanced SE as the normalized power flow in classical photonic simulations,
 which is essentially a model of {\it radiation reaction}.

Recently, however, it was shown that this LDOS
picture of SE breaks down in a medium containing a 
linear amplifier~\cite{franke_fermis_2021,ren_quasinormal_2021}.~
Indeed, it is entirely possible to have a negative LDOS
in a gain medium, even though the medium
is still in the linear regime, which is quantified
by the poles in the Green function (which is only  allowed to have complex poles in the lower complex half plane, i.e., complex eigenfrequency has negative imaginary part).
{The lossy and gain media 
can be modelled with a dielectric function that has positive and negative imaginary parts, respectively.}

Using a quantum field theory, the reason for the breakdown of the LDOS SE formula, 
{with gain}, is related to {the different form of the electric field operator. Indeed, in the presence of gain in a volume $V_{\rm G}$, Eq.~\eqref{eq: Eop_lossy} modifies to 
\begin{align}
    \hat{\bf E}({\bf r},\omega)=&\frac{i}{\omega\epsilon_0}\int_{\mathbb{R}^3-V_{\rm G}} d{\bf r}' {\bf G}({\bf r},{\bf r}',\omega)
\cdot \hat {\bf j}_{\rm L}({\bf r}',\omega)\nonumber\\
&+\frac{i}{\omega\epsilon_0}\int_{V_{\rm G}} d{\bf r}' {\bf G}({\bf r},{\bf r}',\omega)
\cdot \hat {\bf j}_{\rm G}({\bf r}',\omega),
\end{align}
where $\hat {\bf j}_{\rm L}({\bf r},\omega)\propto \sqrt{\epsilon_{\rm Im}^{\rm L}(\mathbf{r},\omega)}\hat {\bf b}({\bf r},\omega)$ is the loss-induced noise operator and $\hat {\bf j}_{\rm G}({\bf r},\omega)\propto \sqrt{\left|\epsilon_{\rm Im}^{\rm G}(\mathbf{r},\omega)\right|}\hat {\bf b}^\dagger({\bf r},\omega)$ is the gain induced noise operator. Consequently,} the fluctuation-dissipation theorem in the form of Eq.~\eqref{eq: SEClass2} gives an additional spatial gain contribution from the operator product $\hat{\mathbf{E}}^\dagger(\mathbf{r}_{\rm 0},\omega)\hat{\mathbf{E}}(\mathbf{r}_{\rm 0},\omega)$.
This results in a nonlocal {\it gain correction} to the SE decay rate,
%
\begin{align}
\begin{split}\label{eq: PF_quan_Gamma}
\Gamma_{\rm quant}^{\rm SE}(\mathbf{r}_{0},\omega)&
= \Gamma^{\rm LDOS}(\mathbf{r}_{0},\omega)+\Gamma^{\rm gain}(\mathbf{r}_{0},\omega),
\\
\end{split}
\end{align}
where
\begin{align}
\label{eq: PF_quan_gain}
\Gamma^{\rm gain}
(\mathbf{r}_{0},\omega)
=\frac{2}{\hbar\epsilon_0}
{\bf d} \cdot 
\mathbf{K}(\mathbf{r}_0,\mathbf{r}_0,\omega)
\cdot {\bf d},
\end{align}
and
\begin{align}
\begin{split}
&\mathbf{K}(\mathbf{r},\mathbf{r}_0,\omega)
=\int_{V_{\rm G}}d\mathbf{s}\Big|\epsilon_{\rm Im}^{\rm G}(\mathbf{s},\omega)\Big|\mathbf{G}(\mathbf{r},\mathbf{s},\omega)\cdot\mathbf{G}^{*}(\mathbf{s},\mathbf{r}_0,\omega),
\end{split}
\label{eq:K}
\end{align}
which 
ensures
that $\Gamma^{\rm SE}_{\rm quant}>0$,
and $\epsilon_{\rm Im}^{\rm G}={\rm Im}[\epsilon^{\rm G}]$ is the imaginary part of the permittivity for the gain medium. 
In addition to the gain modified SE rate, in a quantum derivation,
$\Gamma^{\rm gain}$ leads to excitation from
the ground state to excited state, namely
$\ket{g,{\bf 0}}$ to $\ket{e,{\bf 1}}$.
{We use $|\epsilon_{\rm Im}^{\rm G}|$ to keep positive definite quantities,
and for ease of notation. However, we also highlight that the absolute value appears in the quantum derivation because of the operator ordering associated with gain.}

Related microscopic derivations of gain-modified SE
have also been presented recently~\cite{PhysRevB.105.085116}.
Note that Eq.~\eqref{eq:K} involves a 
{\it nonlocal} contribution from the entire gain medium; in origin, a similar term was added
to the LDOS calculation  to exploit the fluctuation-dissipation theorem, so must be subtracted back off; because of the sign of the imaginary part of the permittivity for gain region, this yields a net positive quantity.
The corresponding quantum Purcell factor is defined as
\begin{align}\label{eq: PF_quan}
F_{\rm P}^{\rm quant}(\mathbf{r}_{0},\omega_{\rm a})=1+\frac{\Gamma^{\rm SE}_{\rm quant}
(\mathbf{r}_{0},\omega_{\rm a})}{\Gamma_{0}(\omega_{\rm a})},
\end{align}
which can be used as a benchmark and a reference result for the ``fixed'' classical results
that we will introduce below.


Although these results are quite general, and can also be described from a quantized mode perspective (using quantized quasinormal modes (QNMs)~\cite{PhysRevA.105.023702}), they cannot be checked for a classical correspondence using simple Maxwell solvers with power flow arguments. Also, without a few-mode description for the Green function, the numerical evaluation of Eq.~\eqref{eq:K} is very difficult and computationally demanding. It is also not known if these results have any classical correspondence.

In this work, 
we present a theory to {\it fix the classical Purcell factor for media containing linear gain amplifiers}, where the same quantum-corrected SE rate can be obtained from purely classical power flow arguments, thus generalizing the usual approach for the nanophotonics
(Maxwell) community, to allow use for media with linear gain. We also give several different viewpoints for this fix, and present simple prescriptions for obtaining this classical correspondence.
Finally, 
using a QNM approach,
we also
show how the classical results are fully obtained from the viewpoint of quantized QNMs in the bad cavity limit. Thus we establish
a classical to quantum picture of SE emission, as well as a quantum to classical picture, in the appropriate limit where this correspondence makes sense (weak coupling).

\section{Classical theory of spontaneous emission
with linear gain media}

The calculation of the classical SE rate
can be obtained from the numerical power flow of a classical dipole, or from the projected LDOS [Eq.~\eqref{eq: Gamma_LDOS}], both of which fail when the medium contains a linear gain amplifier. In this section, we start from the classical power flow from a dipole, and then propose two forms of fixed classical SE rates and Purcell factors with the presence of a linear gain medium. One of these forms is shown to be fully consistent with the quantum results obtained in Ref.~\onlinecite{franke_fermis_2021}.

\subsection{Classical power flow from a polarization dipole}

In a standard Maxwell solver, one can obtain various classical power flows numerically. For example, the power dissipated
 from a point dipole at ${\bf r}_0$, can be computed from a surface integral over the Poynting vector ${\bf S}^{\rm poyn}({\bf r}, \omega)=\frac{1}{2}{\rm Re}\left[\mathbf{E}({\bf r}, \omega)\times\mathbf{H}^{*}({\bf r}, \omega)\right]$ with magnetic field $\mathbf{H}(\mathbf{r},\omega)=\frac{1}{i\omega\mu_0}\nabla\times\mathbf{E}(\mathbf{r},\omega)$~\cite{Novotny2007Principles}, using
\begin{align}
\begin{split}\label{eq: P_spoyn_stru}
P_{\rm }^{\rm }({\bf r}_0,\omega)&=\int_{\Sigma} \hat{{\bf n}} \cdot {\bf S}_{\rm }^{\rm poyn}({\bf r}, \omega) d\mathbf{r},
\end{split}
\end{align}
where the selected surface $\Sigma$ determines the power flow contribution of interest, and the unit vectors, $\hat{{\bf n}}$, are normal to the selected surface, and point outwards.

\begin{figure*}[htb]
    \centering
    \includegraphics[width = 1.99\columnwidth]{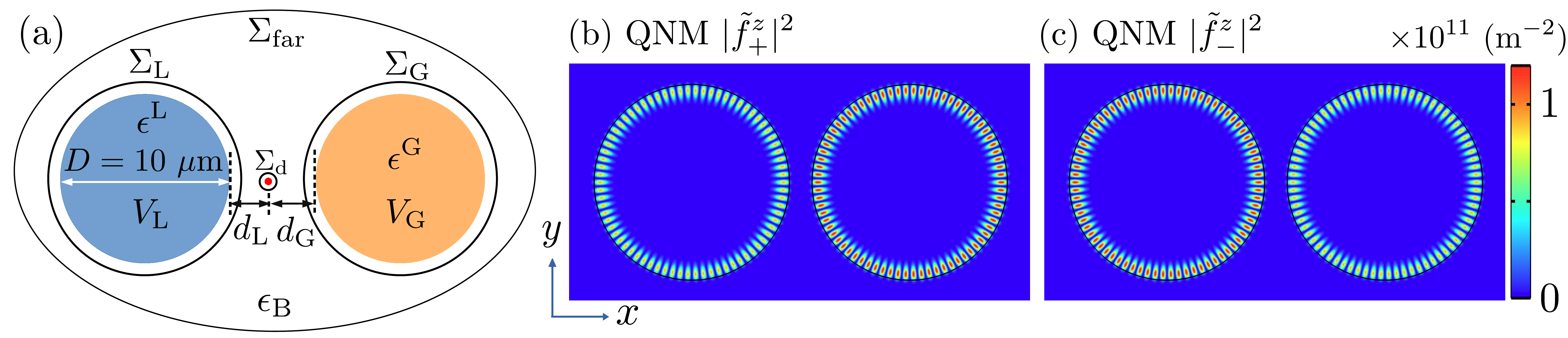}
    \caption{(a)~Schematic diagram of the coupled loss-gain microdisks in free space ($\epsilon_{\rm B}=1$) and the various surfaces ($\Sigma_{\rm d}, \Sigma_{\rm L},\Sigma_{\rm G}, \Sigma_{\rm far}$) and the volume regions ($V_{\rm L}, V_{\rm G}$) for the integrations. Both microdisks have a diameter $D=10~\mu$m, and the permittivities are $\epsilon^{\rm L}={(2+i10^{-5})^2}$ (loss) and $\epsilon^{\rm G}={(2-i5\times10^{-6})^2}$ (gain). The gap distance $d_{\rm gap}=d_{\rm L}+d_{\rm G}$ between the microdisks will be chosen as $1155$~nm and $1160~$nm (close to an exceptional point region), where $d_{\rm L}$ ($d_{\rm G}$) is the minimal distance between the potential point dipole and the lossy (gain) microdisk. 
    $\Sigma_{\rm d}$ is a surface that surrounds only the point dipole. $\Sigma_{\rm L}$ ($\Sigma_{\rm G}$) is a surface that surrounds only the lossy (gain) part of the cavity. $\Sigma_{\rm far}$ is a surface that surrounds the whole coupled-cavity dipole system. $V_{\rm L}$ ($V_{\rm G}$) is the volume for the lossy region and the gain region. Note that when converting the 3D structure to a 2D case, the surface integration will be line integration, and the volume integration will become surface integration.
    (b,c) Coupled QNMs distribution $|\tilde{f}_{\pm}^{z}|^2$ (see text) with $d_{\rm gap}=1160~$nm, where the coupled modes are delocalized around both resonators with a different intensity. 
    }
    \label{fig: sche}
\end{figure*}

Let us consider the example of a dipole emitter located within or near some finite-size inhomogeneous cavity, as shown in Fig.~\ref{fig: sche} (red dot shows the emitter).
In general, there are four kinds of surfaces of interest for electromagnetic power flow from the dipole:
(i) the surface that only encloses the dipole, 
$\Sigma_{\rm d}$, which yields
the local total power flow $P_{\rm LDOS}(\mathbf{r}_{0})$
from the dipole at some location $\mathbf{r}_{0}$ with the presence of the cavity;
(ii) the surface that only encloses 
the lossy part of the cavity system,
$\Sigma_{\rm L}$,
which yields the net positive power $P_{\rm nloss}(\mathbf{r}_{0})$ \textit{flowing into} the lossy region and dissipated within the lossy region, which leads to nonradiative power loss;
(iii) the surface that only encloses the gain part of the cavity,
$\Sigma_{\rm G}$, 
which gives
the net positive power $P_{\rm gain}(\mathbf{r}_{0})$ \textit{flowing out from} the  gain region;
and (iv) the surface that encloses both the dipole and the entire cavity,
$\Sigma_{\rm far}$,
which yields the
outgoing {\it radiative power} $P_{\rm rloss}(\mathbf{r}_{0})$ emitted to the far field region.
For clarity, we distinguish radiative and nonradiative loss with the labels `rloss'
and `nloss', respectively.

These four power flow contributions are  defined from 
\begin{align}
P_{\rm LDOS}(\mathbf{r}_{0},\omega)&=\int_{\Sigma_{\rm d}} \hat{{\bf n}} \cdot {\bf S}_{\rm }^{\rm poyn}({\bf r}, \omega)d\mathbf{r},\label{eq: Ptot_num}\\
P_{\rm nloss/gain}(\mathbf{r}_{0},\omega)&= -{\rm sgn}(\epsilon_{\rm Im}^{\rm L/G}(\omega))
\int_{\Sigma_{\rm L/G}} \hat{{\bf n}} \cdot {\bf S}_{\rm }^{\rm poyn}({\bf r}, \omega)d\mathbf{r},\label{eq: power_nrad_surf_loss}\\
P_{\rm rloss}(\mathbf{r}_{0},\omega)&=\int_{\Sigma_{\rm far}} \hat{{\bf n}} \cdot {\bf S}_{\rm }^{\rm poyn}({\bf r}, \omega)d\mathbf{r},\label{eq: power_rad}
\end{align}
where the sign function in Eq.~\eqref{eq: power_nrad_surf_loss} (${\rm sgn}[\epsilon_{\rm Im}^{\rm L/G}]={\rm sgn}[{\rm Im}(\epsilon^{\rm L/G})]=\pm1$) is used to ensure net positive powers for $P_{\rm nloss/gain}(\mathbf{r}_{0},\omega)$.

The geometry of these surfaces does not have any specific shape requirement, as long as they surround the corresponding sections. 
The two loss/gain regions can also be related to the energy dissipation/amplification in that lossy/gain region, which 
can also be defined in terms of a volume integral~\cite{Novotny2007Principles}:
\begin{align}
\begin{split}
\label{eq: power_nrad_vol_lossgain}
&P_{\rm nloss/gain}(\mathbf{r}_{0},\omega)\\
&={\rm sgn}(\epsilon_{\rm Im}^{\rm L/G}(\omega))\frac{1}{2} \int_{V_{\rm L/G}} {\rm Re}\{\mathbf{J}^{*}_{\rm L/G}(\mathbf{r},\omega)\cdot\mathbf{E}_{\rm L/G}(\mathbf{r},\omega)\}d\mathbf{r},
\end{split}
\end{align}
with the current source within lossy/gain region ({$\mathbf{E}_{\rm L/G}(\mathbf{r},\omega)$ is the electric field within the lossy/gain region})
\begin{align}
\begin{split}
&\mathbf{J}_{\rm L/G}(\mathbf{r},\omega)=-i\omega\epsilon_{0}(\epsilon^{\rm L/G}(\mathbf{r},\omega)-1)\mathbf{E}_{\rm L/G}(\mathbf{r},\omega),
\end{split}
\end{align}
where we note only the imaginary part
$\epsilon_{\rm Im}$ of the permittivity contributes to power loss and gain.

In linear 
media~\cite{Novotny2007Principles}, the net energy flow into (out of) a lossy (gain) region is equal to the energy dissipation (amplification) within this region, which can be obtained from 
\begin{align}\label{eq: surf_vol}
P&=-\int_{\partial V}\mathbf{S}^{\rm poyn}(\mathbf{r},\omega)\cdot\mathbf{\hat{n}}d\mathbf{r}\nonumber\\
&=\frac{1}{2}\int_{V}{\rm Re}\{\mathbf{J}^{*}(\mathbf{r},\omega)\cdot\mathbf{E}(\mathbf{r},\omega)\}d\mathbf{r}.
\end{align}
Thus we can evaluate
$P_{\rm nloss/gain}(\mathbf{r}_{0},\omega)$
using either the surface integral
[Eq.~\eqref{eq: power_nrad_surf_loss}] or the volume integral formalism [Eq.~\eqref{eq: power_nrad_vol_lossgain}].

The  four 
power contributions
satisfy a power conservation rule,
\begin{align}
\begin{split}\label{eq: equiv_power}
&P_{\rm LDOS}(\mathbf{r}_0,\omega)+P^{\rm }_{\rm gain}(\mathbf{r}_0,\omega)
=P_{\rm rloss}(\mathbf{r}_0,\omega)+P^{\rm }_{\rm nloss}(\mathbf{r}_0,\omega),
\end{split}
\end{align}
where either a surface or volume formula can be used.
We can first define a power flow, $P^{\rm SE}$, 
that is related the total SE rate,
as a sum of the far field radiation (raditive loss) and the lossy material nonradiative part:
\begin{align}
    P^{\rm SE}(\mathbf{r}_0,\omega)\equiv P_{\rm rloss}(\mathbf{r}_0,\omega)+P^{\rm }_{\rm nloss}(\mathbf{r}_0,\omega).
\end{align}
Alternatively, using Eq.~\eqref{eq: equiv_power}, we can also define this as
\begin{align}
\label{eq: power_SE2}
P^{\rm SE}(\mathbf{r}_0,\omega)
= P_{\rm LDOS}(\mathbf{r}_0,\omega)+P_{\rm gain}(\mathbf{r}_0,\omega).
\end{align}

We immediately 
recognize that the second form of
$P^{\rm SE}$ [Eq.~\eqref{eq: power_SE2}], which is
in terms of an LDOS term and a gain term,
 is completely analogous to the quantum mechanical 
contributions shown in Eq.~\eqref{eq: PF_quan_Gamma}.
Moreover, 
in a simple dielectric 
structure, then we obtain the usual
$P^{\rm SE}
= P_{\rm LDOS}=P_{\rm rloss}$, with all photons emitted radiatively to the far field.
Below, we will connect these various power flow terms, in a gain-loss medium, with classical decay rates and show a 
clear 
classical-quantum 
correspondence.

\subsection{Classical Purcell factors and decay rates with linear gain based on power flow}

From a practical and simple Maxwell equation viewpoint, one is 
interested in 
the classical dipole-induced power flow that best connects to the SE rates
and Purcell factors.
The classical SE decay rate
is simply
\begin{equation}
\Gamma^{\rm SE}_{\rm class}({\bf r}_0,\omega)
= \Gamma_0(\omega) F^{\rm class}_{\rm P}({\bf r}_0,\omega),
\end{equation}
where $\Gamma_0(\omega)$ is the rate from the point
dipole in the background medium, i.e., without the resonator(s)
or inhomogeneous scattering structure. 

We define
the classical Purcell factors, $F_{\rm P}^{\rm class}({\bf r}_0,\omega)$,
in two different ways,
first from
\begin{align}
\label{eq: FP_mod1}
    F_{\rm P}^{\rm class}(\mathbf{r}_0,\omega)=\frac{P^{\rm SE}({\bf r}_0,\omega)}{P_0(\omega)}=
    \frac{P_{\rm rloss}(\mathbf{r}_0,\omega) + P_{\rm nloss}(\mathbf{r}_0,\omega)}{P_0({\omega})},   
\end{align}
or, alternatively, from
\begin{align}
\label{eq: FP_mod2}
    F_{\rm P}^{\rm class}(\mathbf{r}_0,\omega)=\frac{P^{\rm SE}({\bf r}_0,\omega)}{P_0(\omega)}=
     \frac{P_{\rm LDOS}(\mathbf{r}_0,\omega) + P_{\rm gain}(\mathbf{r}_0,\omega)}{P_0({\omega})},  
\end{align}
where $P_0(\omega)$ is the power flow from the point
dipole in the background medium, i.e., without the resonator(s)
or inhomogenous scattering structure. 
In contrast, when using only the contribution from the LDOS, then 
\begin{equation}
F_{\rm P}^{\rm LDOS}(\mathbf{r}_0,\omega)=
\frac{P_{\rm LDOS}(\mathbf{r}_0,\omega)}{P_0({\omega})}.
\end{equation}

Thus from Eq.~\eqref{eq: FP_mod2}, we find the following relationship for the SE decay rate:
\begin{align}
\begin{split}\label{eq: FP_mod2_decay}
    \Gamma^{\rm SE}_{\rm class}(\mathbf{r}_0,\omega)=\Gamma^{\rm LDOS}(\mathbf{r}_0,\omega)+ \Gamma^{\rm gain}_{\rm class}(\mathbf{r}_0,\omega),
\end{split}    
\end{align}
with 
\begin{equation}
\Gamma^{\rm LDOS}(\mathbf{r}_0,\omega)=
\Gamma_0(\omega)\frac{P_{\rm LDOS}(\mathbf{r}_0,\omega)}{P_0(\omega)},
\end{equation}
and
\begin{align}\label{eq: gamma_gain_class}
    \Gamma^{\rm gain}_{\rm class}(\mathbf{r}_0,\omega)=\Gamma_0(\omega)\frac{P_{\rm gain}(\mathbf{r}_0,\omega)}{P_0(\omega)},
\end{align}
where the rates are obtained from the normalized classical power flows.~
Equation~(\ref{eq: FP_mod2_decay}) is in an identical form to the quantum derivation, Eq.~\eqref{eq: PF_quan_Gamma}. 
Similarly, we can define 
the total SE decay rates in terms of the  radiative (far field emission), and nonradiation decay within the lossy region, using
\begin{align}
\begin{split}\label{eq: FP_mod1_decay}
    \Gamma^{\rm SE}_{\rm class}(\mathbf{r}_0,\omega)=\Gamma^{\rm rloss}_{\rm class}(\mathbf{r}_0,\omega)+ \Gamma^{\rm nloss}_{\rm class}(\mathbf{r}_0,\omega),
\end{split}    
\end{align}
where
\begin{align}
\Gamma^{\rm rloss}_{\rm class}(\mathbf{r}_0,\omega)&=\Gamma_0(\omega)\frac{P_{\rm rloss}(\mathbf{r}_0,\omega)}{P_0(\omega)},\label{eq: gamma_far_class}
\end{align}
and
\begin{align}
\Gamma^{\rm nloss}_{\rm class}(\mathbf{r}_0,\omega)&=\Gamma_0(\omega)\frac{P_{\rm nloss}(\mathbf{r}_0,\omega)}{P_0(\omega)}\label{eq: gamma_loss_class}. 
\end{align}

Equations~\eqref{eq: FP_mod1_decay} and \eqref{eq: FP_mod2_decay} demonstrate that 
one can associate
SE decay from 
the 
 far field radiative decay plus the
 nonradiative decay from the lossy region; or, as in the quantum result,
 from  the 
 usual LDOS contribution (which may be negative) plus a nonlocal correction from the gain region.
The former is perhaps more appealing, and is valid even for gain media; it also avoids the LDOS picture and does not require a nonlocal gain calculation.

\subsection{Green function solution}\label{sec: Greenfunction}

Next, we focus on the first expression for the classical decay rate $\Gamma_{\rm class}^{\rm SE}$  shown in Eq.~\eqref{eq: FP_mod2_decay}, which has a similar form to the quantum result $\Gamma^{\rm SE}_{\rm quant}$, Eq.~\eqref{eq: PF_quan_Gamma}.
To compare the classical result with the quantum result, one needs to make a clearer connection between $\Gamma^{\rm gain}_{\rm class}$ [Eq.~\eqref{eq: gamma_gain_class}] and $\Gamma^{\rm gain}$ [Eq.~\eqref{eq: PF_quan_gain}], where the latter is expressed in terms of the Green functions. So here we will rewrite $\Gamma^{\rm gain}_{\rm class}$ in terms of 
the Green functions as well. 

The gain induced power, $P_{\rm gain}(\mathbf{r}_0,\omega)$ in $\Gamma^{\rm gain}_{\rm class}$~
can be obtained from a volume 
integration [Eq.~\eqref{eq: power_nrad_vol_lossgain}].~ 
%
At any spatial point, 
the dipole-induced field is 
\begin{align}
\mathbf{E}(\mathbf{r},\omega)=\mathbf{G}(\mathbf{r},\mathbf{r}_{0},\omega)\cdot\frac{\mathbf{d}}{\epsilon_{0}}=\mathbf{G}(\mathbf{r},\mathbf{r}_{0},\omega)\cdot{\mathbf{n}}_{\rm d}\frac{|\mathbf{d}|}{\epsilon_{0}},
\end{align}
where $\mathbf{d}={\mathbf{n}}_{\rm d}|\mathbf{d}|$ with unit vector $\mathbf{{n}}_{\rm d}$.
Therefore,
we can rewrite the
gain contribution to the power flow as 
\begin{align}\label{eq: power_nrad_vol_gain}
\begin{split}
P_{\rm res}^{\rm G}(\mathbf{r}_{0},\omega)
&=-\frac{1}{2} \int_{V_{\rm G}} {\rm Re}\{\mathbf{J}^{*}_{\rm G}(\mathbf{r},\omega)\cdot\mathbf{E}_{\rm G}(\mathbf{r},\omega)\}d\mathbf{r},\\
&=\frac{1}{2} \int_{V_{\rm G}} \omega\epsilon_{0}|\epsilon_{\rm Im}^{\rm G}(\mathbf{r},\omega)||\mathbf{E}_{\rm G}(\mathbf{r},\omega)|^2d\mathbf{r},\\
&=\frac{\omega |\mathbf{d}|^2}{2\epsilon_{0}}\int_{V_{\rm G}}|\epsilon_{\rm Im}^{\rm G}(\mathbf{r},\omega)||\mathbf{G}_{\rm }(\mathbf{r},\mathbf{r}_{0},\omega)\cdot{\mathbf{n}}_{\rm d}|^2d\mathbf{r},
\end{split}
\end{align}
which is the same as $P_{\rm gain}$ defined in Eq.~\eqref{eq: power_nrad_vol_lossgain}, but now 
evaluated with a Green function solution.

Next, we can write the gain contribution to the
{\it classical} decay rate~\cite{Novotny2007Principles,jun_nonresonant_2008,liu_excitation_2009} as
\begin{align}
\begin{split}\label{eq: Gamma_Gvol}
\Gamma^{\rm gain}_{\rm class}(\mathbf{r}_0,\omega)
&=\Gamma_0(\omega)\frac{P_{\rm res}^{\rm G}(\mathbf{r}_{0},\omega)}{P_0(\omega)}=
\frac{P_{\rm res}^{\rm G}(\mathbf{r}_{0},\omega)}{(\hbar\omega)/4}\\
&=\frac{2|\mathbf{d}|^2}{\hbar\epsilon_{0}}\int_{V_{\rm G}}|\epsilon_{\rm Im}^{\rm G}(\mathbf{r},\omega)||\mathbf{G}_{\rm }(\mathbf{r},\mathbf{r}_{0},\omega)\cdot\mathbf{n}_{\rm d}|^2d\mathbf{r},
\end{split}
\end{align}
which can easily be shown to be {\it identical} to $\Gamma^{\rm gain}(\mathbf{r}_{0}, \omega_{\rm a})$ [Eq.~\eqref{eq: PF_quan_gain}] derived in the quantum theory, when $\omega=\omega_{\rm a}$. 
Note that the quantum result is at the resonance frequency $\omega_{\rm a}$ of the atom, and the classical result is at the linear frequency $\omega$ of interest, but it is clear $\omega=\omega_{\rm a}$ when comparing the two.
Thus, we conclude that the classical result  for the total SE decay rate shown in  Eq.~\eqref{eq: FP_mod2_decay} is identical to the quantum result, defined in Eq.~\eqref{eq: PF_quan_Gamma}, so there is indeed a 
classical-quantum correspondence for the Purcell factors as well:
$F_{\rm P}^{\rm class}(\omega=\omega_{\rm a}) = F_{\rm P}^{\rm quant}(\omega_{\rm a})$ [Eq.~\eqref{eq: FP_mod2} and Eq.~\eqref{eq: PF_quan}]. In both cases, one cannot use the usual LDOS formula for SE when any gain is included in the medium description, and there is a non-local correction from the gain medium.

Note that a factor of 4 is needed
when converting dipole-induced power to 
a SE rate with a classical dipole simulation~\cite{Barnes2020}.
However, we highlight that
such a factor is not
needed in a self-consistent semiclassical Maxwell-Bloch solver~\cite{PhysRevA.96.023859,PhysRevA.95.063853}; thus radiation reaction, without any noise, does yield the correct
SE decay rate when seeded with 
classical coherence.

Similarly,
one also has the non-radiative loss rate
\begin{align}
\begin{split}\label{eq: Gamma_Lvol}
\Gamma^{\rm nloss}_{\rm class}(\mathbf{r}_0,\omega)
=\frac{2|\mathbf{d}|^2}{\hbar\epsilon_{0}}\int_{V_{\rm L}}\epsilon_{\rm Im}^{\rm L}(\mathbf{r},\omega)|\mathbf{G}_{\rm }(\mathbf{r},\mathbf{r}_{0},\omega)\cdot\mathbf{n}_{\rm d}|^2d\mathbf{r},
\end{split}
\end{align}
with an explicit volume integration, in terms of the material Green function.

\subsection{Quasinormal mode expansions}\label{sec: QNMs}
In the previous subsection, the gain contribution $\Gamma_{\rm class}^{\rm gain}$ to the SE rates was written in terms of the Green function. However, the 
two-space-point Green function is not easy to compute in general, and 
it is better to obtain these semi-analytically
from an accurate mode theory.
Moreover, often with SE studies, one is interested in practical cavity structures, where enhancements are mainly caused by resonant modes. Indeed, this is precisely the spirit of Purcell's formula, in that it is defined in terms of modal quantities. Thus it is desirable to connect the above general results to structures that can be described in terms of the underlying cavity modes.

Quasinormal modes, $\tilde{\mathbf{f}}_{\mu}$~\cite{lai_time-independent_1990,leung_completeness_1994,leung_time-independent_1994,leung_completeness_1996,lee_dyadic_1999,kristensen_generalized_2012,kristensen_modes_2014,PhysRevX.7.021035,lalanne_light_2018,kristensen_modeling_2020}, are the natural modes of open cavities, which are the solutions to the vector Helmholtz equation,
\begin{equation}\label{eq: helmholtz}
\boldsymbol{\nabla}\times\boldsymbol{\nabla}\times\tilde{\mathbf{f}}_{{\mu}}\left(\mathbf{r}\right)-\left(\dfrac{\tilde{\omega}_{{\mu}}}{c}\right)^{2}
\epsilon(\mathbf{r},\tilde{\omega}_{\mu})\,\tilde{\mathbf{f}}_{{\mu}}\left(\mathbf{r}\right)=0,
\end{equation}
with the Silver-M\"uller radiation condition~\cite{Kristensen2015}. The corresponding eigenfrequencies $\tilde{\omega}_{\mu}=\omega_{\mu}-i\gamma_{\mu}$ are complex, which also
yields the modal quality factor $Q_\mu=\omega_{\mu}/(2\gamma_{\mu})$. 
{To numerically obtain the QNM eigenfunctions and eigenfrequencies, 
we employ an efficient dipole technique in complex frequency space~\cite{bai_efficient_2013-1}. We emphasize that using this technique results in properly normalized QNMs.}

For coupled resonator systems, one can directly find the coupled QNMs by setting $\epsilon(\mathbf{r},\tilde{\omega}_{\mu})$ to be the permittivity for the whole systems.
However, one could also use an accurate coupled QNM theory
~\cite{ren_quasinormal_2021,franke_fermis_2021,ren_connecting_2022,PhysRevA.105.023702,vial_coupling_2016,Kristensen_coupled_modes_2017,tao_coupling_2020}
to get the properties of the hybrid QNMs based on the QNMs of individual resonators, which are also well defined with gain media ~\cite{ren_quasinormal_2021,franke_fermis_2021}.
If one considers two separate cavities described by permittivity $\epsilon^{1(2)}$ in background medium with $\epsilon_{\rm B}$ (as we will show later), each with one QNM of interest $\tilde{\mathbf{f}}_{1/2}$ and corresponding eigenfrequencies $\tilde{\omega}_{1/2}$, the eigenfrequencies for the coupled system will be 
\begin{equation}
\label{eq: omega_pm}
\tilde \omega_{\pm}=
\frac{\tilde\omega_1+\tilde\omega_2}{2}
\pm \frac{\sqrt{4\tilde \kappa_{12}\tilde \kappa_{21} + (\tilde\omega_1-\tilde\omega_2)^2}}{2},
\end{equation}
with the coupling coefficients $\tilde \kappa_{12/21}$ ($i,j=1,2$)
\begin{equation}\label{eq: kappa_1221}
\tilde \kappa_{ij}=
\frac{\tilde \omega_j}{2} \int_{V_{i}} d{\bf r} [\epsilon^i({\bf r})-\epsilon_{\rm B}] \tilde {\bf f}_i({\bf r}) \tilde {\bf f}_j({\bf r}).
\end{equation}
The coupled QNMs are 
\begin{align}\label{eq: QNMs_pm}
\ket{\tilde{\bf f}_\pm}&=
\frac{\tilde\omega_\pm-\tilde\omega_2}{\sqrt{(\tilde\omega_\pm-\tilde\omega_2)^2+ \tilde \kappa_{21}^2}}
\ket{\tilde{\bf f}_1}
\nonumber \\
&
+ \frac{-\tilde\kappa_{21}}{\sqrt{(\tilde\omega_\pm-\tilde\omega_2)^2+ \tilde \kappa_{21}^2}} 
\ket{\tilde{\bf f}_2}.
\end{align}
For the example shown later, the notation $1,2$ here will be replaced by $\rm L,G$.

Once the two hybrid QNMs $\tilde{\mathbf{f}}_{\pm}$ are obtained, in the frequency regime of interest (a total of two QNMs dominate, which we have checked to be accurate for the studies below), the photon Green function near or within the resonators can be obtained from 
a QNM expansion~\cite{leung_completeness_1994,ge_quasinormal_2014},
\begin{align}\label{eq: GFwithpm}
\begin{split}
\mathbf{G}\left(\mathbf{r},\mathbf{r}_{0},\omega\right)&=\sum_{\mu}A_{\mu}(\omega)\tilde{\mathbf{f}}_{\mu}\left({\bf r}\right)\tilde{\mathbf{f}}_{\mu}\left({\bf r}_{0}\right)\\
&\approx A_{+}(\omega)\tilde{\mathbf{f}}_{+}\left({\bf r}\right)\tilde{\mathbf{f}}_{+}\left({\bf r}_{0}\right)+A_{-}\left(\omega\right)\tilde{\mathbf{f}}_{-}({\bf r})\tilde{\mathbf{f}}_{-}\left({\bf r}_{0}\right),
\end{split}
\end{align}
where $A_{\pm}(\omega)=\omega/[2(\tilde{\omega}_{\pm}-\omega)]$.

Next, one can rewrite the gain contribution $\Gamma_{\rm class}^{\rm gain}$ to the SE rates with the volume integration form, in terms of the QNMs:\\

\begin{align}
\label{eq: Gamma_Gvol_QNM}
\Gamma_{\rm QNM}^{\rm gain}(\mathbf{r}_0,\omega)
&=\frac{2|\mathbf{d}|^2}{\hbar\epsilon_{0}}\int_{V_{\rm G}}|\epsilon_{\rm Im}^{\rm G}(\mathbf{r},\omega)| \nonumber
\\
&  \times \left|\left(\sum_{\mu} A_{\mu}\left(\omega\right)\,\tilde{\mathbf{f}}_{\mu}\left({\bf r}\right)\tilde{\mathbf{f}}_{\mu}\left({\bf r}_{0}\right)\right)\cdot\mathbf{n}_{\rm d}\right|^2d\mathbf{r},
\end{align}
which we stress again is identical to the quantum result, Eq.~\eqref{eq: PF_quan_gain}, when applying the same QNM expansion and $\omega=\omega_{\rm a}$.
Note that
 two QNMs $\tilde{\mathbf{f}}_{\pm}$ are included in $\sum_{\mu}$, though one can increase this if required, so we keep this more general form.

Similarly, 
one can  also write $\Gamma^{\rm nloss}_{\rm class}$ in terms of QNMs, through
\begin{align}
\label{eq: Gamma_Lvol_QNM}
\Gamma_{\rm QNM}^{\rm nloss}(\mathbf{r}_0,\omega)
&=\frac{2|\mathbf{d}|^2}{\hbar\epsilon_{0}}\int_{V_{\rm L}}\epsilon_{\rm Im}^{\rm L}(\mathbf{r},\omega) \nonumber
\\
&  \times\left|\left(\sum_{\mu} A_{\mu}\left(\omega\right)\,\tilde{\mathbf{f}}_{\mu}\left({\bf r}\right)\tilde{\mathbf{f}}_{\mu}\left({\bf r}_{0}\right)\right)\cdot\mathbf{n}_{\rm d}\right|^2d\mathbf{r}.
\end{align}

Note that one could also rewrite the gain (and loss) contributions 
to the SE rates with the surface integration form, in terms of QNMs.
However, since the QNMs are obtained outside the resonator, this is not as accurate as using the QNMs within a volume integral~\cite{ge_quasinormal_2014,franke_impact_2023}, so we keep the volume integral form above.

Using Eqs.~\eqref{eq: FP_mod2} and \eqref{eq: FP_mod2_decay}, we now define the {\it corrected classical Purcell factors}
in terms of the QNMs, as 
\begin{align}\label{eq: FP_cQNM_volsurf_2}
  F_{\rm P,QNM}^{\rm class}(\mathbf{r}_0,\omega)= F_{\rm P,QNM}^{\rm LDOS}(\mathbf{r}_0,\omega)+\frac{\Gamma^{\rm gain}_{\rm QNM}(\mathbf{r}_0,\omega)}{\Gamma_{0}(\omega)},
\end{align}
where the LDOS contribution is the usual Purcell formula, 
\begin{align}\label{eq: FP_QNM_LDOS}
F_{\rm P,QNM}^{\rm LDOS}(\mathbf{r}_0,\omega)=1+\frac{\mathbf{d}\cdot{\rm Im}[\mathbf{G}^{\rm QNM}(\mathbf{r}_{0},\mathbf{r}_{0},\omega)]\cdot\mathbf{d}}{\mathbf{d}\cdot{\rm Im}[\mathbf{G}_{\rm hom}(\omega)]\cdot\mathbf{d}},
\end{align}
with 
\begin{align}
&\mathbf{G}^{\rm QNM}(\mathbf{r}_{0},\mathbf{r}_{0},\omega)
\nonumber \\
&=
A_{+}\left(\omega\right)\tilde{\mathbf{f}}_{+}\left({\bf r}_0\right)\tilde{\mathbf{f}}_{+}\left({\bf r}_{0}\right)+A_{-}\left(\omega\right)\tilde{\mathbf{f}}_{-}\left({\bf r}_0\right)\tilde{\mathbf{f}}_{-}\left({\bf r}_{0}\right).
\end{align}
In the limit of a few QNMs,
these LDOS and gain factors are trivial to compute,
and in our numerical example below, we will show excellent agreement (versus full-dipole numerical simulations) using just two QNMs.

 For the numerical example w  consider later, the far field radiative contribution is negligible, thus by using Eqs.~\eqref{eq: FP_mod1} and \eqref{eq: FP_mod1_decay}, we can define the {\it corrected classical Purcell factors}
in terms of the QNMs, with
\begin{align}\label{eq: FP_cQNM_volsurf_1}
  F_{\rm P,QNM}^{\rm class}(\mathbf{r}_0,\omega)\approx 1+\frac{\Gamma^{\rm nloss}_{\rm QNM}(\mathbf{r}_0,\omega)}{\Gamma_{0}(\omega)},
\end{align}
where $\Gamma^{\rm nloss}_{\rm QNM}(\mathbf{r}_0,\omega)$ is given from Eq.~\eqref{eq: Gamma_Lvol_QNM}. 

To summarize the key classical equations, we have presented {\it fixed} classical Purcell 
formulas that can easily be applied to completely arbitrary loss-gain systems with no mode approximations, 
by using Eqs.~\eqref{eq: FP_mod1}-\eqref{eq: FP_mod2}, or one can exploit QNMs for cavities
using Eq.~\eqref{eq: FP_cQNM_volsurf_2}.
The latter is also appealing as it can connect to more rigorous approaches using quantized QNMs,
which can then be checked in the bad cavity limit~\cite{franke_quantization_2019,PhysRevA.105.023702}, {namely when one can adiabatically eliminate the cavity modes
and derive the modified SE rate 
in a semiclassical Purcell regime}.

\subsection{Full-dipole numerical solutions}\label{sec: fulldipole}

Naturally, one also desires to compute the
SE rates without any insight from a
Green function solution.
Since the dipole power flow can in principle be obtained from any classical Maxwell equation solver, one can compute the numerical classical Purcell factors $F_{\rm P}^{\rm num,1/2}$, from 
\begin{align}
\begin{split}\label{eq: FP_num_mod1}
    F_{\rm P}^{\rm num,1}({\bf r}_0, \omega) 
    = \frac{P_{\rm rloss}(\mathbf{r}_0,\omega) + P_{\rm nloss}(\mathbf{r}_0,\omega)}{P_0({\mathbf{r}_0,\omega})},
\end{split}    
\end{align}
or
\begin{align}
\begin{split}\label{eq: FP_num_mod2}
    F_{\rm P} ^{\rm num,2}({\bf r}_0, \omega) 
    = \frac{P_{\rm LDOS}(\mathbf{r}_0,\omega) + P_{\rm gain}(\mathbf{r}_0,\omega)}{P_0({\mathbf{r}_0,\omega})},
\end{split}    
\end{align}
with the numerical power 
\begin{equation}
P_0(\mathbf{r}_{0},\omega)=\int_{\Sigma_{\rm d}} \hat{{\bf n}} \cdot {\bf S}_{\rm background}^{\rm }({\bf r}, \omega) d\mathbf{r},
\end{equation}
from the point
dipole in the background medium. The term, $F_{\rm P}^{\rm num,1/2}$, can be compared with the quantum Purcell factors $F_{\rm P}^{\rm quant}$ [Eq.~\eqref{eq: PF_quan}] from the corrected Fermi's golden rule, and $F_{\rm P,QNM}^{\rm class}$~[Eq.~\eqref{eq: FP_cQNM_volsurf_2}] from QNMs.
Note that  we simply use the labels
`1' and `2' for the purpose of showing numerical results later, but mathematically these should yield identical results. So one can use either form.

Note for the specific example we will consider below, we could use the approximated form $F_{\rm P}^{\rm num,1}({\bf r}_0, \omega)\approx1+P_{\rm nloss}/P_0$, as the nonradiative part is dominating.

In contrast, without using any mode expansions, 
 the numerically
exact Purcell factor contribution from the LDOS can be obtained from  
\begin{align}
\begin{split}\label{eq: FP_num_LDOS}
    F_{\rm P,num} ^{\rm LDOS}({\bf r}_0, \omega) 
    = \frac{P_{\rm LDOS}(\mathbf{r}_0,\omega)}{P_0({\mathbf{r}_0,\omega})}.
\end{split}    
\end{align}

\section{Numerical Results for Coupled Gain-Loss Resonators}\label{sec: loss_gain}

\begin{figure*}
    \centering
    \includegraphics[width = 0.79 \columnwidth]{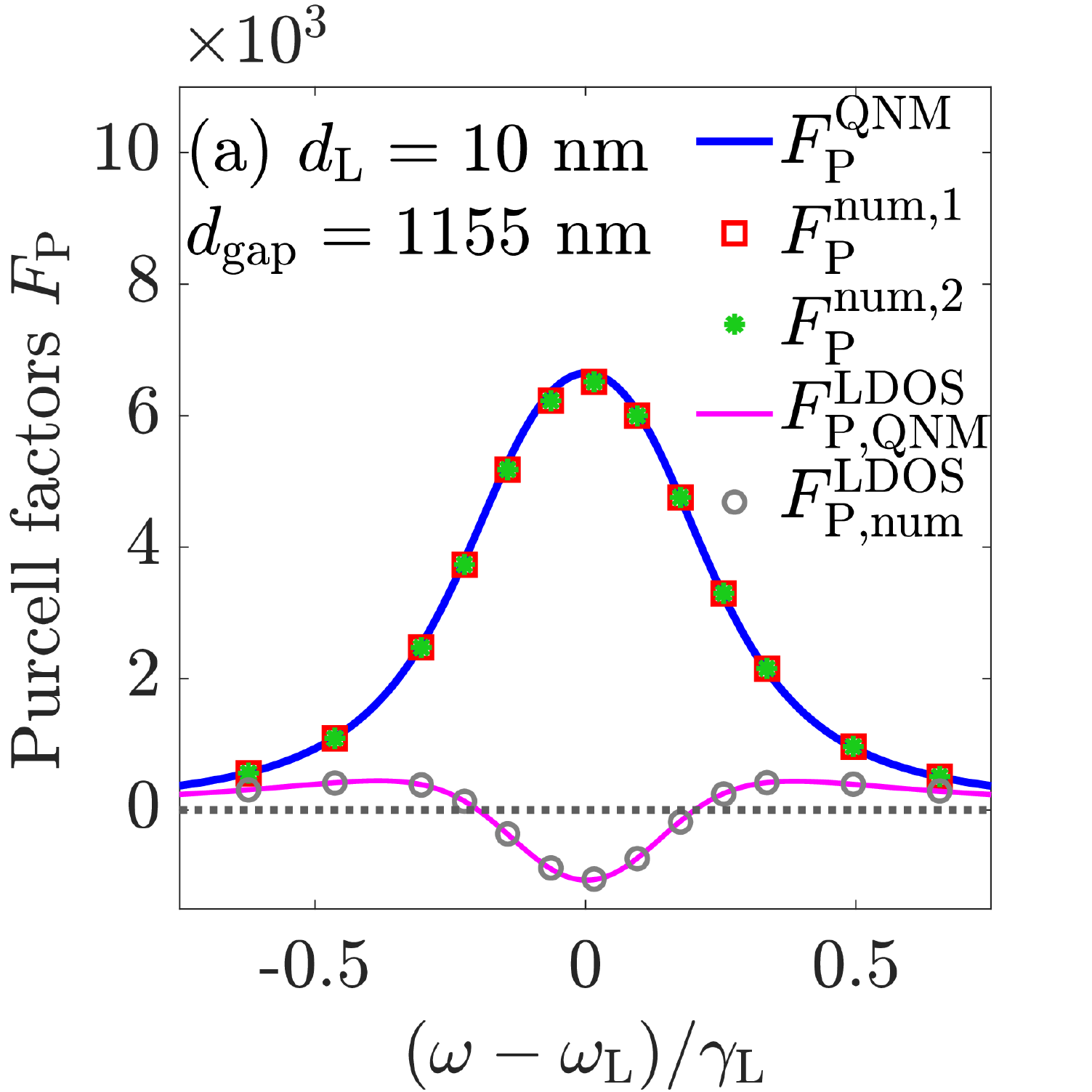}    
    \includegraphics[width = 0.79 \columnwidth]{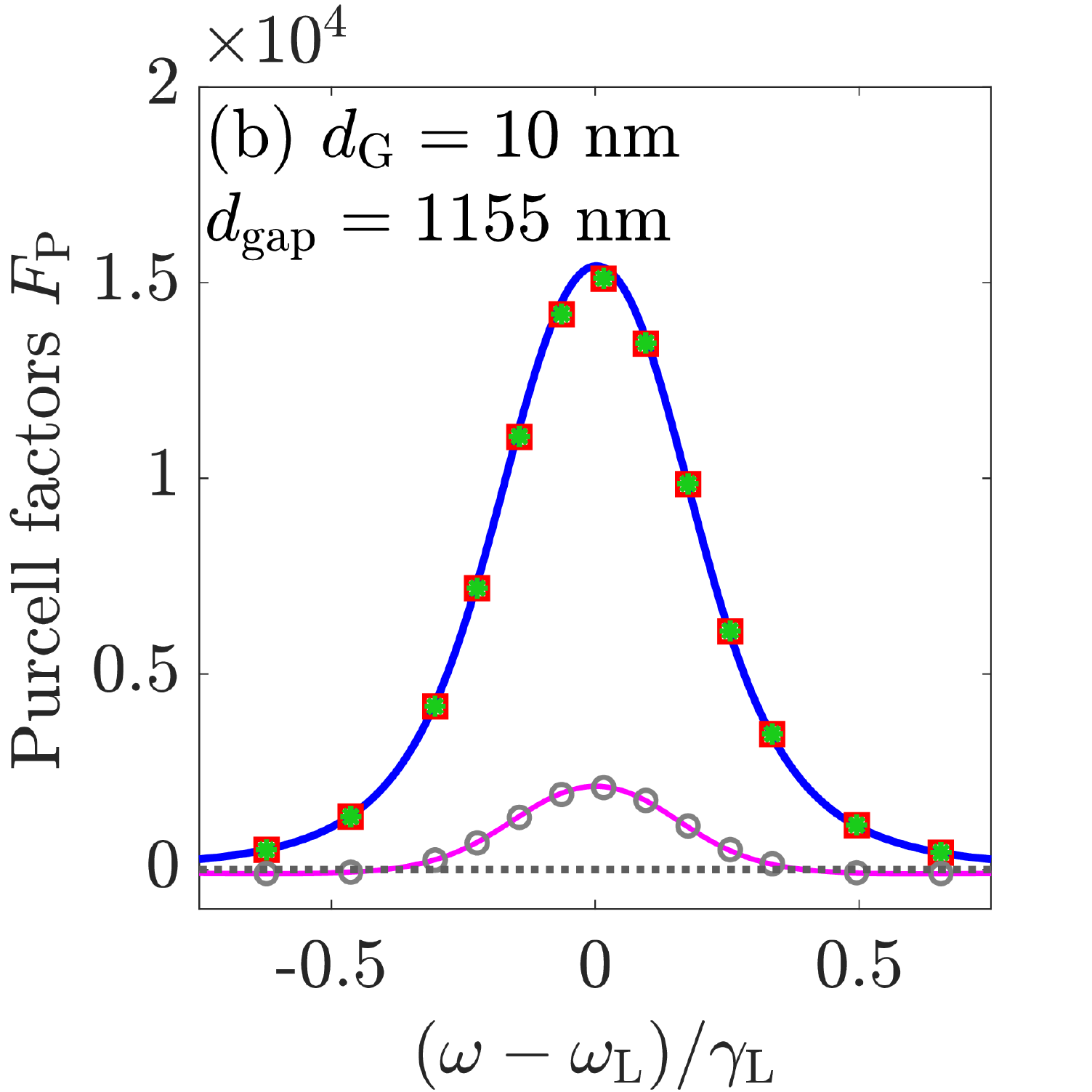}    
    \includegraphics[width = 0.79 \columnwidth]{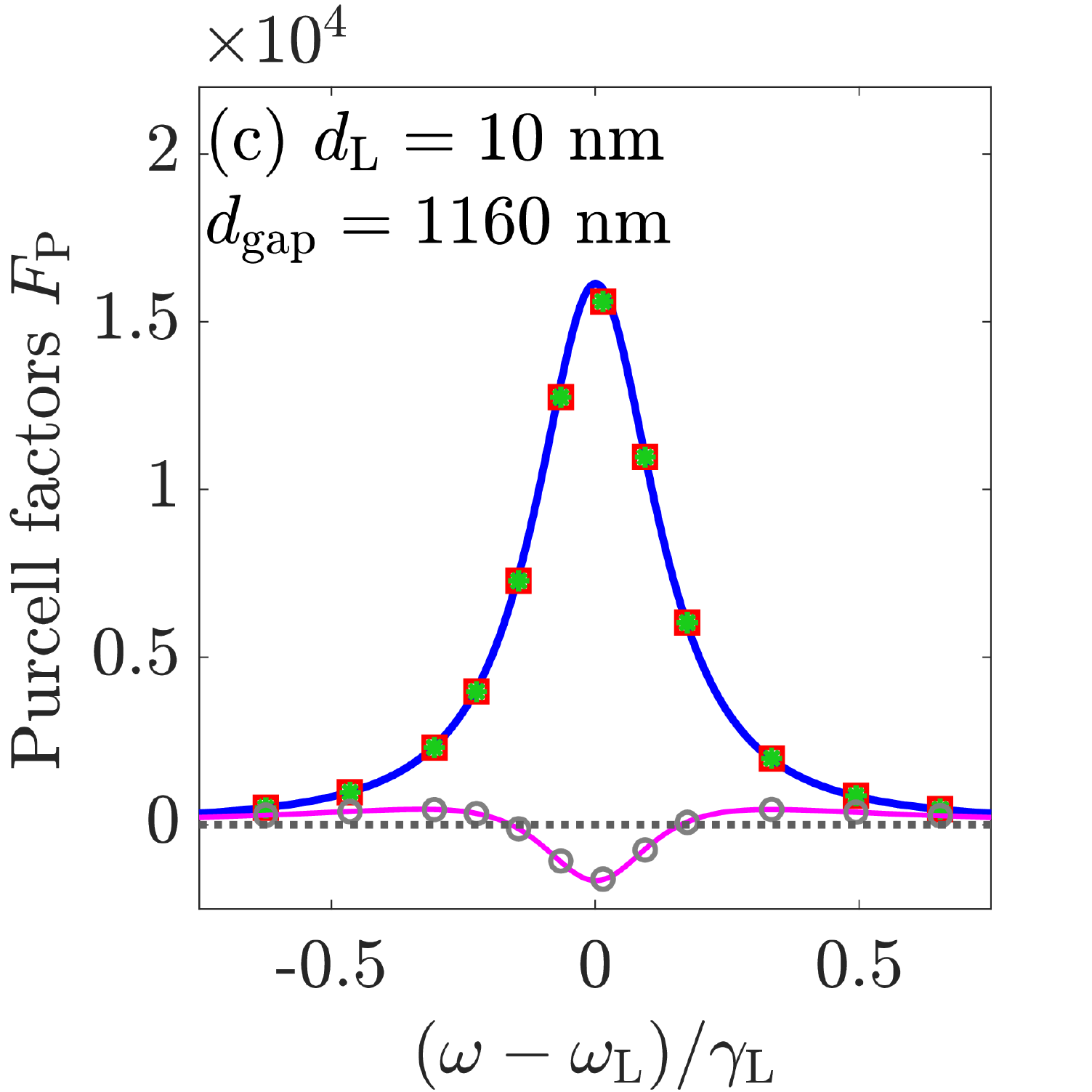}    
    \includegraphics[width = 0.79 \columnwidth]{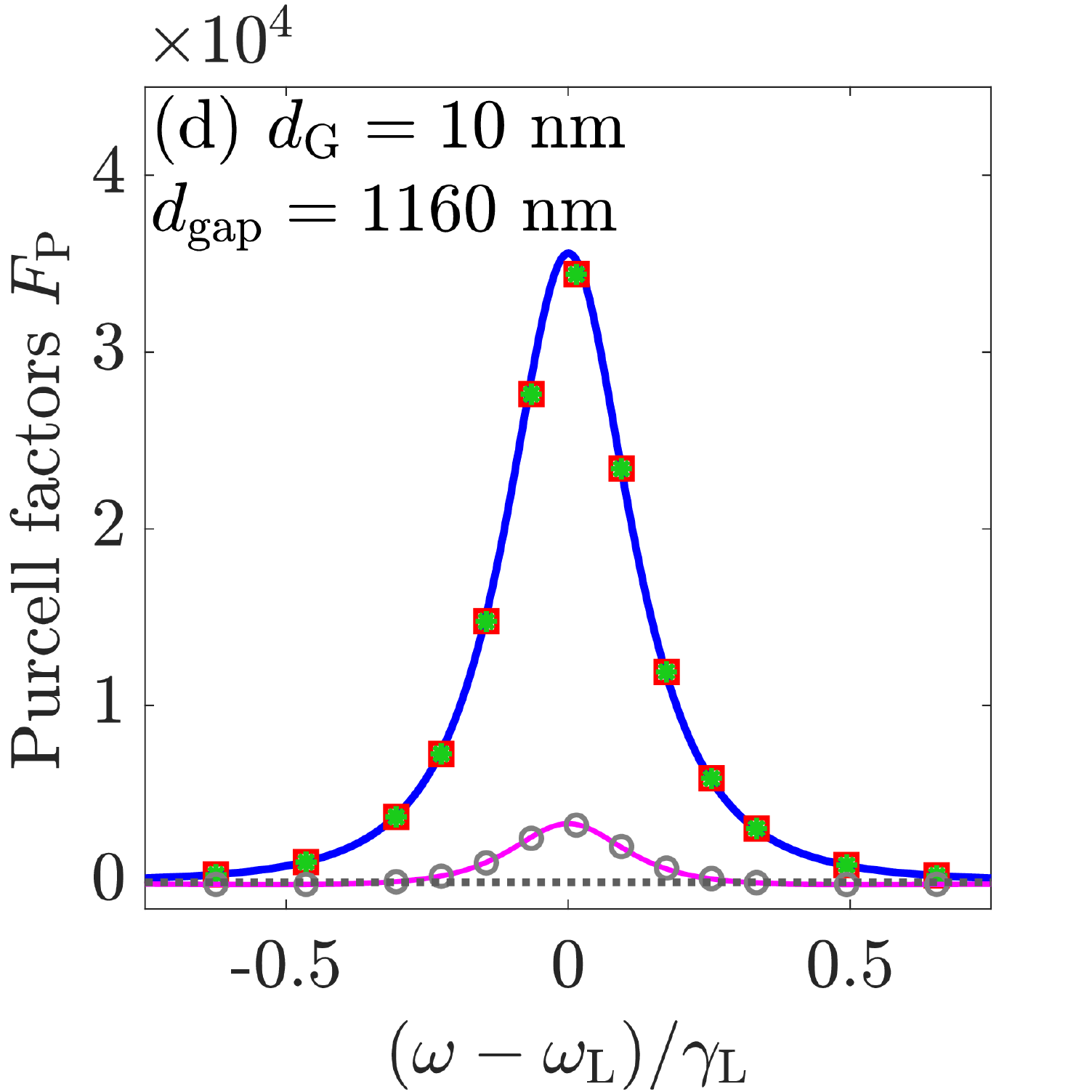}    
   \caption{
    Corrected numerical Purcell factors $F_{\rm P}^{\rm num,1/2}$ [Eqs.~\eqref{eq: FP_num_mod1} and \eqref{eq: FP_num_mod2}, red squares and green asterisks], agreeing very well with the Purcell factors using a two-QNM expansion $F_{\rm P}^{\rm QNM}=F_{\rm P,QNM}^{\rm class}=F_{\rm P}^{\rm quant}$ (solid blue curves); the same answer is obtained classically [Eq.~\eqref{eq: FP_cQNM_volsurf_2} or Eq.~\eqref{eq: FP_cQNM_volsurf_1}] and quantum mechanically [Eq.~\eqref{eq: PF_quan}]. 
    For comparison, the usual results from just the LDOS are shown, using the numerical solution $F^{\rm LDOS}_{\rm P,num}$ [Eq.~\eqref{eq: FP_num_LDOS}, grey circles] and the QNM solution $F^{\rm LDOS}_{\rm P,QNM}$ [Eq.~\eqref{eq: FP_QNM_LDOS}, magenta curve]. The dotted grey curve indicates the value of $0$. Note that $\hbar\omega_{\rm L}\approx 0.83~{\rm eV}$ and $\hbar\gamma_{\rm L}\approx 4~\mu{\rm eV}$, which are related to the real part and imaginary part of the eigenfrequency, $\tilde{\omega}_{\rm L}=\omega_{\rm L}-i\gamma_{\rm L}$, for the single QNM $\mathbf{\tilde{f}}_{\rm L}$ of lossy cavity only. 
    In (a-b), the rates are evaluated for the gap distance $d_{\rm gap}=1155~$nm, with the dipole at (a) $d_{\rm L}=10~$nm (close to the lossy resonator) and (b) $d_{\rm G}=10~$nm (close to the gain resonator). (c-d)
    Same as in (a-b), but with 
    $d_{\rm gap}=1160~$nm.
    In all cases, we stress there are no fitting parameters, and we obtain negative values for the LDOS rates in a certain frequency range.
    }
    \label{fig: FP}
\end{figure*}

We next present numerical examples, using a coupled loss-gain resonator system, similar to those previously studied in Refs.~\cite{franke_fermis_2021,ren_quasinormal_2021}. The system consists of two 2D microdisks with permittivities $\epsilon^{\rm L}=(2+i10^{-5})^2$ (loss) and $\epsilon^{\rm G}=(2-i5\times10^{-6})^2$ (gain), inside a homogeneous medium with $\epsilon_{\rm B}=1$ as shown in Fig.~\ref{fig: sche}(a). Each microdisk has a diameter $D=10~\mu$m. The dipole is placed within the gap, which is $d_{\rm L}$($d_{\rm G}$) away from the lossy (gain) resonator and $d_{\rm L}+d_{\rm G}=d_{\rm gap}$ is satisfied, where the gap distance $d_{\rm gap}=1160~$nm or $1155~$nm (close to the exceptional point region~\cite{ren_quasinormal_2021}, where the two resonances approach) and either $d_{\rm L}=10~$nm (close to the loss cavity) or $d_{\rm G}=10~$nm (close to the gain cavity) is selected. 

We first calculate the dominant single QNM $\tilde{\mathbf{f}}_{\rm L}$ for the 
loss microdisk in the frequency regime of interest (single mode approximation), which is a TM mode $(\tilde{h}_x, \tilde{h}_y, \tilde{f}_z)$ (the magnetic QNM $\tilde{\mathbf{h}}$  is polarized in the $xy$ plane, and the electric QNM $\tilde{\mathbf{f}}$ only has a $z$ component) with radial mode number $q=1$, and azimuthal mode number $m=37$. The mode eigenfrequency is $\tilde{\omega}_{\rm L}=\omega_{\rm L}-i\gamma_{\rm L}=1.266666\times10^{15}-i6.26\times10^{9}~({\rm rad}/{\rm s})$ ($\hbar\tilde{\omega}_{\rm L}=\hbar\omega_{\rm L}-i\hbar\gamma_{\rm L}\approx 0.83 {\rm eV}-i4~\mu{\rm eV}$) with a quality factor around $Q_{\rm L}=\omega_{\rm L}/(2\gamma_{\rm L})\approx10^{5}$ (resonant wavelength around $\lambda_{\rm L}=2\pi c/\omega_{\rm L}\approx1487$~nm). 
We also obtain the QNM $\tilde{\mathbf{f}}_{\rm G}$ for the gain microdisk 
in the same frequency regime of interest. The corresponding eigenfreqeuncy is $\tilde{\omega}_{\rm G}=\omega_{\rm G}-i\gamma_{\rm G}\sim\omega_{\rm L}+i0.5\gamma_{\rm L}$.

Then we use coupled QNM theory~\cite{ren_quasinormal_2021,franke_fermis_2021,ren_connecting_2022,PhysRevA.105.023702}, to efficiently obtain the eigenfrequencies $\tilde{\omega}_{\pm}$ [Eq.~\eqref{eq: omega_pm}] and the QNMs $\tilde{\mathbf{f}}_{\pm}$ [Eq.~\eqref{eq: QNMs_pm}] for the hybridized QNMs in an analytical form. For example, when the gap distance $d_{\rm gap}=1160~$nm, the distribution for the 
$z$-components of the coupled QNMs are shown in Fig.~\ref{fig: sche} (b,c), which are located in both cavities with different intensities.
After checking the validity of using the QNMs 
to model the resonator response, the Green functions can be obtained from a QNM expansion,
 as described in Eq.~\eqref{eq: GFwithpm}.


We will consider four test cases with the following gap distances and dipole locations: $d_{\rm gap}=1155~$nm or $d_{\rm gap}=1160~$nm and $d_{\rm L}=10~$nm or $d_{\rm G}=10~$nm. First, for comparison, we show the results with the LDOS contribution only in Fig.~\ref{fig: FP}, where the grey circles  show the full numerical dipole results $F_{\rm P,num}^{\rm LDOS}$ [Eq.~\eqref{eq: FP_num_LDOS}] and the solid magenta curve shows the QNMs results $F_{\rm P,QNM}^{\rm LDOS}$ 
[Eq.~\eqref{eq: FP_QNM_LDOS}], which agree quantitatively well with each other. However, with the contribution from the LDOS only, the SE rates are greatly underestimated and can also be negative.

As shown in Ref.~\cite{franke_fermis_2021} and in Eqs.~\eqref{eq: PF_quan_Gamma},~\eqref{eq: PF_quan_gain}, and \eqref{eq: PF_quan}, the corrected Fermi's golden rule with the linear gain will yield a net-positive Purcell factor when adding a gain contribution to the LDOS contribution.
On the other hand, in this work, starting from the classical power flow, we have also derived the same results (see 
Eqs.~\eqref{eq: FP_mod2_decay}, \eqref{eq: Gamma_Gvol} and \eqref{eq: Gamma_Gvol_QNM} with a volume integration form) as the corrected Fermi's golden rule, i.e., $F_{\rm P,QNM}^{\rm class}(\omega=\omega_{\rm a})=F_{\rm P}^{\rm quant}(\omega_{\rm a})$ (Eq.~\eqref{eq: FP_cQNM_volsurf_2} and Eq.~\eqref{eq: PF_quan} are identical analytically) when using the same QNMs expansion of the Green function. 
For all four cases, the (corrected) classical Purcell factors are net-positive (see solid blue curves). In addition, especially for the examples studied here (where the far field decay is negligible), the alternative approximate form Eq.~\eqref{eq: FP_cQNM_volsurf_1} gives basically the same results as the solid blue curves.

Importantly, these fixed Purcell factors ($F_{\rm P}^{\rm quant}=F_{\rm P,QNM}^{\rm class}=F_{\rm P}^{\rm QNM}$) show excellent agreement with full dipole solutions $F_{\rm P}^{\rm num,1/2}$ (Eqs.~\eqref{eq: FP_num_mod1} and \eqref{eq: FP_num_mod2}, red squares and green asterisks), as seen from Fig.~\ref{fig: FP}.~ 
This indicates the validity of the classical Purcell factors and SE rates defined in  Eqs.~\eqref{eq: FP_mod2}, \eqref{eq: FP_mod2_decay}, \eqref{eq: FP_mod1}, and \eqref{eq: FP_mod1_decay}, where the first two tell us that an additional net-positive gain contribution should be added to the general LDOS contribution to account for the total SE rates in the case with linear gain; this agrees with the corrected Fermi's golden rule~\cite{franke_fermis_2021}, and was verified to have the same analytical expression when using the Green function solution and a volume integration form. 
Moreover, in this work, the alternative forms Eq.~\eqref{eq: FP_mod1}, and Eq.~\eqref{eq: FP_mod1_decay} show that the far field radiative and the nonradiative part within the lossy region also account for the total SE rates as well, without having to use the  LDOS contribution at all. This alternative form is possibly more appealing, and it works in linear media with both gain and lossy parts;
in addition, this latter form is more convenient for also defining the radiative $\beta$ factors, from: $\beta_{\rm rad} = P_{\rm rloss}/(P_{\rm rloss}+P_{\rm nloss})$. This quantity is always less than 1, as expected in a linear medium.
In contrast, clearly, one cannot use
$\beta_{\rm rad} = P_{\rm rloss}/P_{\rm LDOS}$, unless the medium is lossy only.

{
\section{Accounting for local-field effects for quantum emitters embedded inside the loss or gain materials}

In many real situations, the emitter is embedded within the loss or gain resonator 
material~\cite{Kumar2005}.
This leads to a well-known 
local-field problem of a finite-size emitter, since the LDOS
is divergent for a point dipole~\cite{PhysRevA.60.4094,VanVlack2012,PhysRevA.74.023803,Ge2015}. 
For high-index structures, the 
{\it real cavity model} is the most appropriate, where one models the finite-size emitter as a small cavity with real permittivity
$\epsilon_{\rm c}$, with a resonant point dipole in the center. 

For resonant nanophotonic structures, it is still possible to extract the dominant QNM contribution to the local-field
SE rate, which may also contain background
contributions that stem simply from the
small cavity in a background medium~\cite{Ge2015}.
In this section, we will show that
our QNM theory above works also with local-field corrections, without any change in 
formalism, if one computes the QNMs in the presence of the local-field cavity.
We will show examples when the dipole
is embedded within a lossy disk or gain disk resonator.

\begin{figure*}[htb]
    \centering
    \includegraphics[width = 1.69\columnwidth]{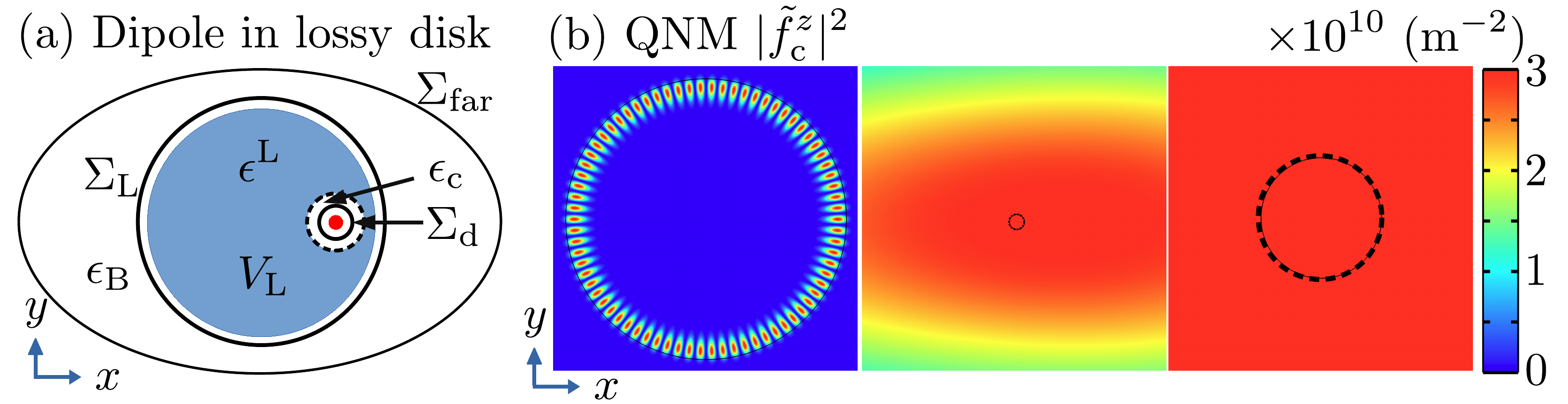}    
    \includegraphics[width = 0.89\columnwidth]{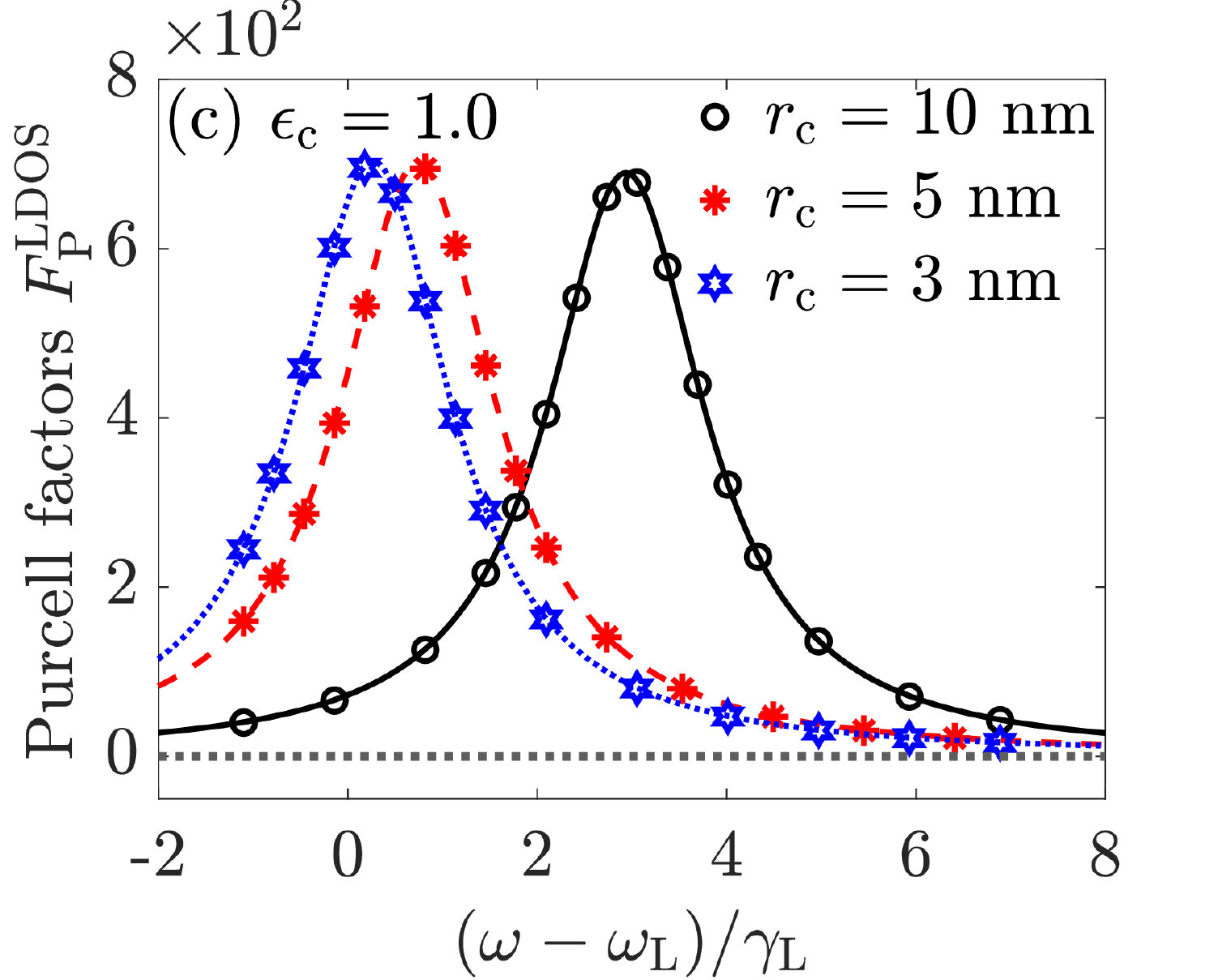}
    \includegraphics[width = 0.89\columnwidth]{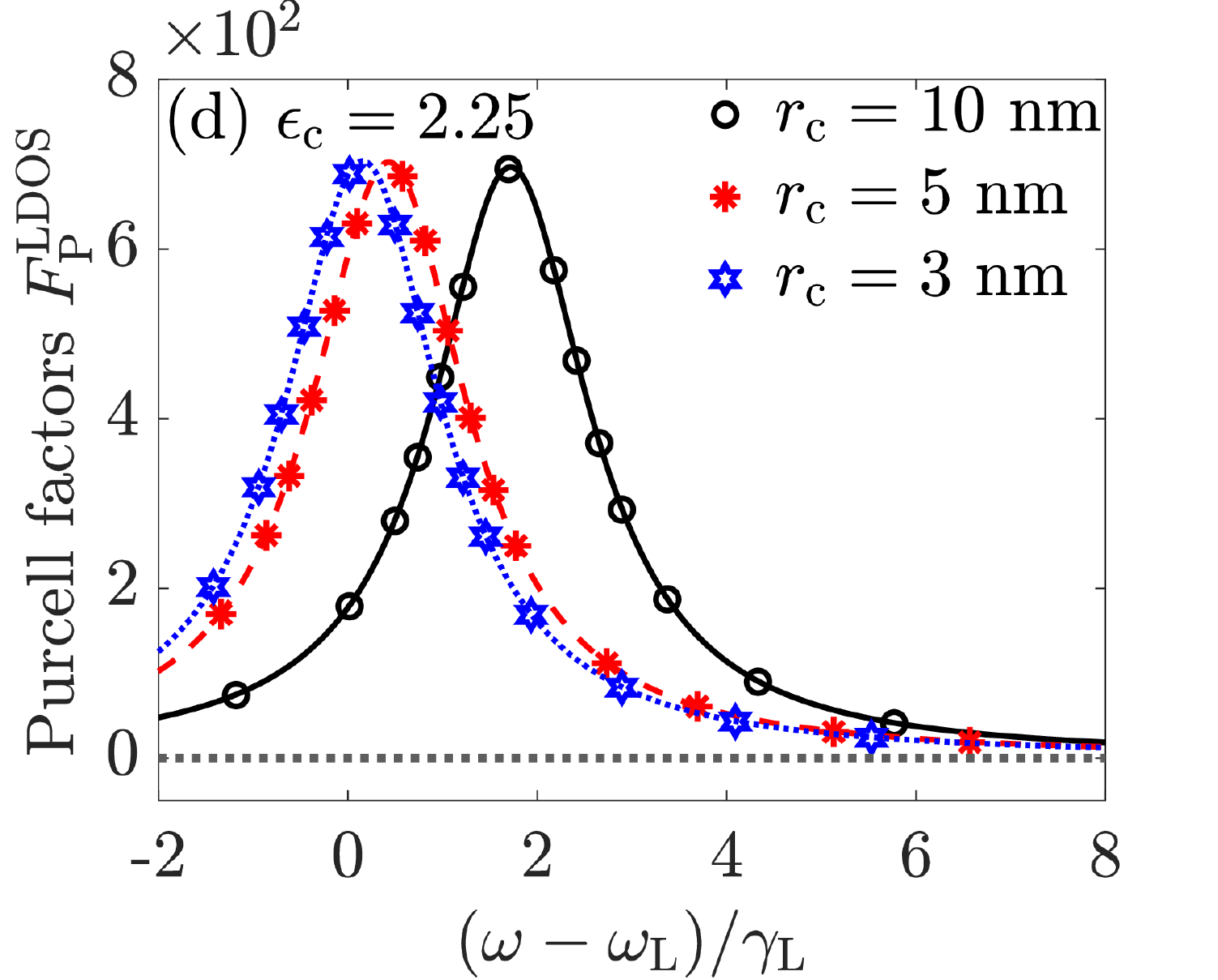}    
    \caption{{(a)~Schematic of a single lossy microdisk [with diameter $D=10~\mu$m and permittivity $\epsilon^{\rm L}=(2+i10^{-5})^2$] in free space ($\epsilon_{\rm B}=1$). Within the lossy disk, is a small cavity region (labelled by a dashed circle,  much smaller compared with the size of the disk, not to scale) which has a radius, $r_{\rm c}$, and a real permittivity $\epsilon_{\rm c}$. The smallest distance between the center of this dashed circle and the surface of the disk is $330~$nm.  Various surfaces ($\Sigma_{\rm d}, \Sigma_{\rm L}, \Sigma_{\rm far}$) and the volume region ($V_{\rm L}$) for the integration are shown. (Note that now the surface $\Sigma_{\rm L}$ encloses both the lossy region and the point dipole, so when calculating $P_{\rm nloss}$, for simplicity, we will only use the volume form as shown in Eq.~\eqref{eq: power_nrad_vol_lossgain}.)
    (b) An example dominant single QNM distribution $|\tilde{f}_{\rm c}^{z}|^2$ (and zoom-in) is shown, in the case of $r_{\rm c}=5~$nm and $\epsilon_{\rm c}=1.0$. The zoom-in plots show the 
    field distribution in the vicinity of the dashed circle region (real cavity). (c) Classical Purcell factors of a dipole placed at the center [see the red dot in (a)] of the dashed circle with various sizes (black/red/blue: $r_{\rm c}=10/5/3~$nm) while the permittivity is fixed at $\epsilon_{\rm c}=1.0$. We see very good agreements for all three cases between classical QNMs ($F_{\rm P,QNM}^{\rm LDOS}(\mathbf{r}_0,\omega)$,~curves, Eq.~\eqref{eq: FP_QNM_LDOS_loc_onlyloss}) and full numerical dipole results ($F_{\rm P,num} ^{\rm LDOS}({\bf r}_0, \omega)$,~markers, Eq.~\eqref{eq: FP_num_LDOS}). In addition, the red shift of the resonance is found with the decrease in the radius of the dashed circle. (d) Similar to (c), but with fixed permittivity $\epsilon_{\rm c}=2.25$. Once again, we see excellent agreement between classical QNMs and full dipole results, and a red shift of the resonance when decreasing $r_{\rm c}$ is found. Moreover, with the same $r_{\rm c}$, a red shift of the resonance is also found when permittivity $\epsilon_{\rm c}$ is increased, by comparing (c) and (d).}
    }
    \label{fig: sche_loc_onlyloss}
\end{figure*}

\begin{figure*}[htb]
    \centering
    \includegraphics[width = 1.99\columnwidth]{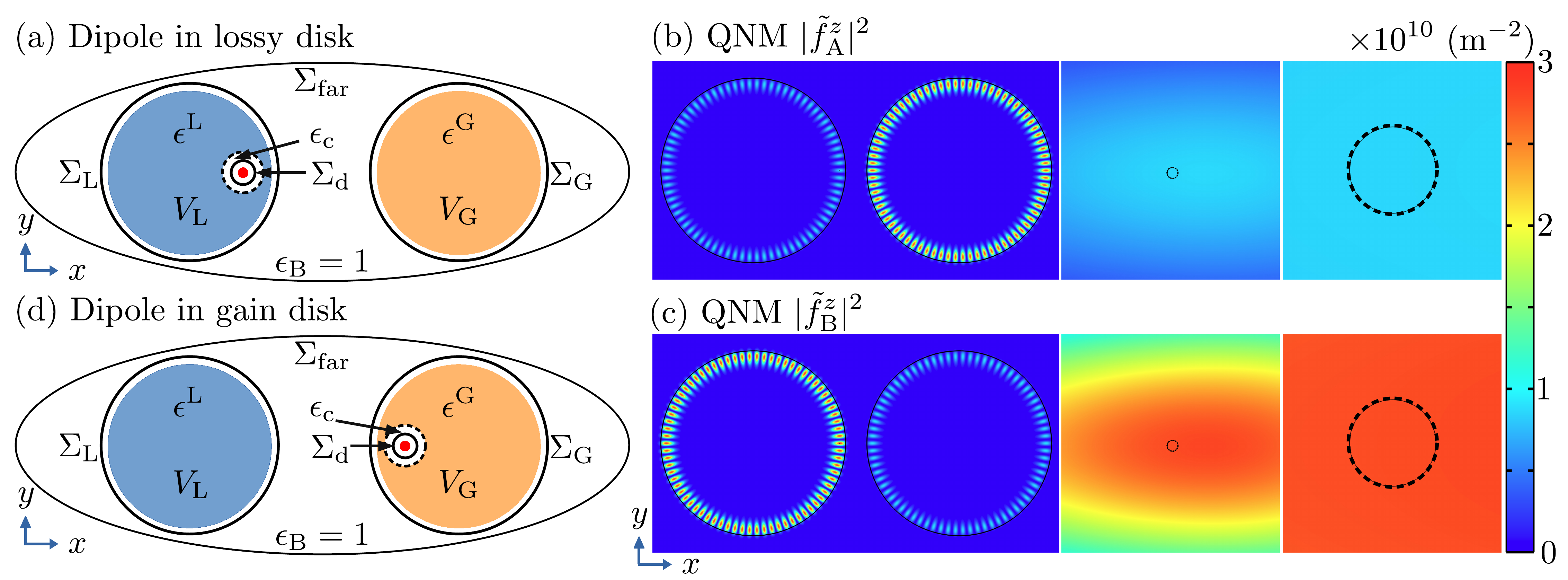}   
    \caption{{(a) Schematic of the coupled loss-gain microdisks, which is similar to the one shown in Fig.~\ref{fig: sche_loc_onlyloss}(a), but (i) the gain disk is added back with gap distance $d_{\rm gap}=1155~$nm; and (ii) the permittivity of the gain disk is $\epsilon^{\rm G}=(2-i1\times10^{-6})^2$ [or $\epsilon^{\rm G}=(2-i2\times10^{-6})^2$], which is different from $\epsilon^{\rm G}=(2-i5\times10^{-6})^2$ that we used in previous sections (Fig.~\ref{fig: sche} and Fig.~\ref{fig: FP}). We decrease the amount of gain to make sure the coupled QNMs are having positive $\gamma_{\rm A/B}$ (with $\tilde{\omega}_{\rm A/B}=\omega_{\rm A/B}-i\gamma_{\rm A/B}$). Note that the surface $\Sigma_{\rm L}$ encloses both the lossy region and the point dipole, so when calculating $P_{\rm nloss}$, for simplicity, we will only use the volume form as shown in Eq.~\eqref{eq: power_nrad_vol_lossgain}.
    (b,c) The field distribution of two dominant QNMs $|\tilde{f}_{\rm A/B}^{z}|^2$,  with zoom-in plots, when the small dashed circle is located within the lossy disk [Fig.~\ref{fig: sche_loc_lossgain} (a)], where $r_{\rm c}=5~$nm and $\epsilon_{\rm c}=1.0$. The permittivity of the gain disk is set as $\epsilon^{\rm G}=(2-i1\times10^{-6})^2$.
    (d) Similar to (a), but now a point dipole is placed within the small dashed circle (real cavity), that is inside the gain disk. Here again, for simplicity, we only use the volume form Eq.~\eqref{eq: power_nrad_vol_lossgain} to get $P_{\rm gain}$ since the surface $\Sigma_{\rm G}$ encloses both the gain region and the point dipole.}
    }
    \label{fig: sche_loc_lossgain}
\end{figure*}

\subsection{Single lossy resonator microdisk}

First, we focus on the case with a single lossy microdisk, since this alone is a non-trivial 
and an important problem. As shown in Fig.~\ref{fig: sche_loc_onlyloss}(a), 
we consider a circle region, with a radius of $r_{\rm c}$ (labelled by the dashed circle, not to scale), where the permittivity is $\epsilon_{\rm c}$ (real).
This circle represents the surface of the real cavity, namely the finite size emitter.
The line formed by the center of this circle and the center of the lossy disk is parallel to the $x-$axis. The shortest distance between the center of the circle and the surface of the disk is $330~$nm. The point dipole 
(labelled by the red dot) is placed within this circle region (the solid black circle surrounding the dipole is $\Sigma_{\rm d}$, which is used to calculate LDOS power in the full dipole numerical method).

In the presence of this small circle (cavity) with permittivity $\epsilon_{\rm c}$, a dominant single QNM is found in the frequency range of interest. For example, with $\epsilon_{\rm c}=1.0$ and $r_{\rm c}=5~$nm, the dominant single QNM distribution $|\tilde{f}^{z}_{\rm c}|^2$ is shown in Fig.~\ref{fig: sche_loc_onlyloss}(b), where the zoom-in region close to the small circle is also shown.

The LDOS Purcell factors with no gain materials (i.e., the LDOS Purcell factors are identical to the total Purcell factors) are given by
\begin{align}\label{eq: FP_QNM_LDOS_loc_onlyloss}
F_{\rm P,QNM}^{\rm LDOS}(\mathbf{r}_0,\omega)=1+\frac{\mathbf{d}\cdot{\rm Im}[\mathbf{G}^{\rm QNM}(\mathbf{r}_{0},\mathbf{r}_{0},\omega)]\cdot\mathbf{d}}{\mathbf{d}\cdot{\rm Im}[\mathbf{G}_{\rm hom}(\omega)]\cdot\mathbf{d}}
\end{align}
where 
\begin{align}
\mathbf{G}^{\rm QNM}(\mathbf{r}_{0},\mathbf{r}_{0},\omega)\approx A_{\rm c}(\omega)\tilde{\mathbf{f}}_{\rm c}(\mathbf{r}_{0})\tilde{\mathbf{f}}_{\rm c}(\mathbf{r}_{0}),
\end{align}
and
$\tilde{\mathbf{f}}_{\rm c}$ and $\tilde{\omega}_{\rm c}$ are the dominant single QNM and the corresponding 
eigenfrequency.
The QNM expansion coefficient is defined similar to before, i.e.,
$A_{\rm c}(\omega)=\omega/[2(\tilde{\omega}_{\rm c}-\omega)]$. 

For the point dipole within the small cavity circle, the results are shown in Fig.~\ref{fig: sche_loc_onlyloss}(c) and (d), where we confirm the excellent agreement between the QNMs results $F_{\rm P,QNM}^{\rm LDOS}(\mathbf{r}_0,\omega)$ (curves, Eq.~\eqref{eq: FP_QNM_LDOS_loc_onlyloss}) and full numerical dipole methods $F_{\rm P,num} ^{\rm LDOS}({\bf r}_0, \omega)$ (markers, Eq.~\eqref{eq: FP_num_LDOS}). Note the power $P_{\rm LDOS}(\mathbf{r}_0,\omega)$ is obtained via surface $\Sigma_{\rm d}$ shown in Fig.~\ref{fig: sche_loc_onlyloss} (a). Moreover, as shown  Fig.~\ref{fig: sche_loc_onlyloss}(c), a red shift of the resonance is found when one decreases the size of the 
real cavity while keeping a fixed permittivity $\epsilon_{\rm c}=1.0$. A similar phenomenon is found when the permittivity is fixed at $\epsilon_{\rm c}=2.25$ as shown in Fig.~\ref{fig: sche_loc_onlyloss} (d). Furthermore, a red shift of the resonance is also obtained when one increases permittivity from $\epsilon_{\rm c}=1.0$ to $\epsilon_{\rm c}=2.25$, while keeping the same size of the circular region, by comparing Fig.~\ref{fig: sche_loc_onlyloss}(c) and (d).

Although we have modelled the dipoles and disks in 2D, the general expressions will also work for 3D 
geometries as well.  In the 3D, there may also be a more significant impact from non-QNM contributions, which can typically be also included analytically~\cite{Ge2015}. However, in our examples, this contribution is clearly negligible and not needed so that everything can be accurately captured from only the QNM
scattering contribution. Having everything in terms of a QNM, including the real cavity, is extremely convenient and numerically efficient, which also applies in the case of QNM quantization.

\subsection{Coupled loss-gain disks}

Next, we will consider our main system of interest using coupled loss-gain disks, where the small real cavity (circular region) is now within the lossy disk 
[Fig.~\ref{fig: sche_loc_lossgain}(a)] or within the gain disk [Fig.~\ref{fig: sche_loc_lossgain} (d)]. Similar to the above subsection (loss resonator only), the line formed by the center of the circle and the center of the loss/gain disks is parallel to the $x$-axis. The smallest distance between the center of the circle and the surface of the disk
is $330~$nm. The radius of the real-cavity circle is fixed at $r_{\rm c}=5~$nm, and the permittivity is fixed at $\epsilon_{\rm c}=1.0$ in this subsection.

The diameter of the two disks and the permittivity of the lossy disks are again $D=10~\mu$m and $\epsilon^{\rm L}=(2+i10^{-5})^2$, as in Fig.~\ref{fig: sche} and Fig.~\ref{fig: FP}. The gap distance is fixed at $d_{\rm gap}=1155~$nm. However, we will use $\epsilon^{\rm G}=(2-i1\times10^{-6})^2$ or $\epsilon^{\rm G}=(2-i2\times10^{-6})^2$ [which is different from $\epsilon^{\rm G}=(2-i5\times10^{-6})^2$ that we used in Sec.~\ref{sec: loss_gain}], in order to ensure the Green functions have poles in the lower complex half plane (as required for a linear medium).

With the dipole excitation technique~\cite{bai_efficient_2013-1}, two coupled QNMs (QNM A and QNM B) are found directly in the frequency region of interest, now including the real cavity of interest. To keep things clear, we do not employ the coupled QNM theory to obtain the hybrid QNMs for the examples in this subsection.
As an example, when the circular region is within the lossy disk [Fig.~\ref{fig: sche_loc_lossgain} (a)] and the permittivity of the gain disk is $\epsilon^{\rm G}=(2-i1\times10^{-6})^2$, the distribution of two QNMs $|\tilde{f}^{z}_{\rm A/B}|^2$ are shown in Fig.~\ref{fig: sche_loc_lossgain}(b) and (c), which live in both disks. The zoom-in region close to the circle is also shown, where the field is distributed smoothly.

\begin{figure*}[htb]
    \centering
    \includegraphics[width = 0.89\columnwidth]{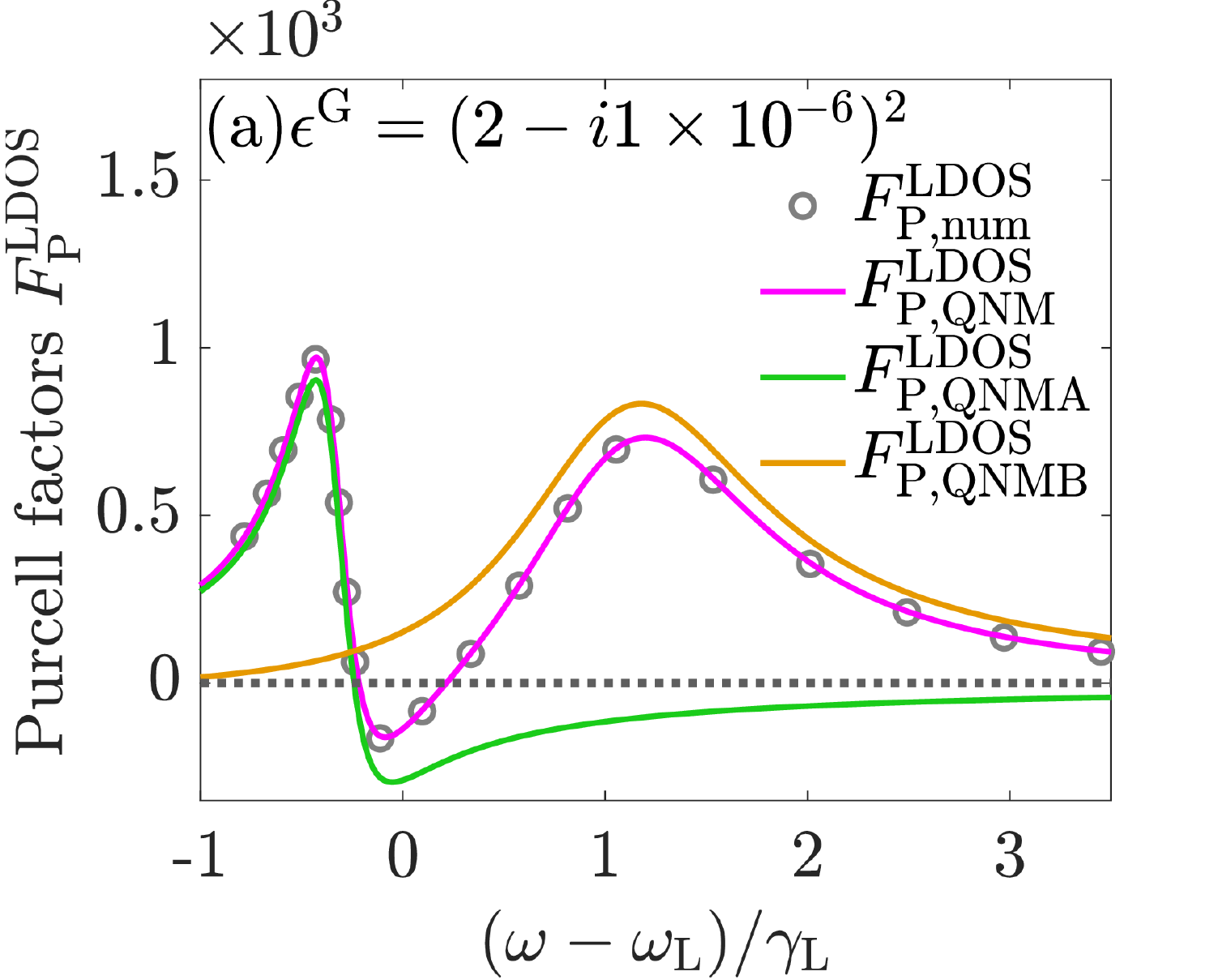}
    \includegraphics[width = 0.89\columnwidth]{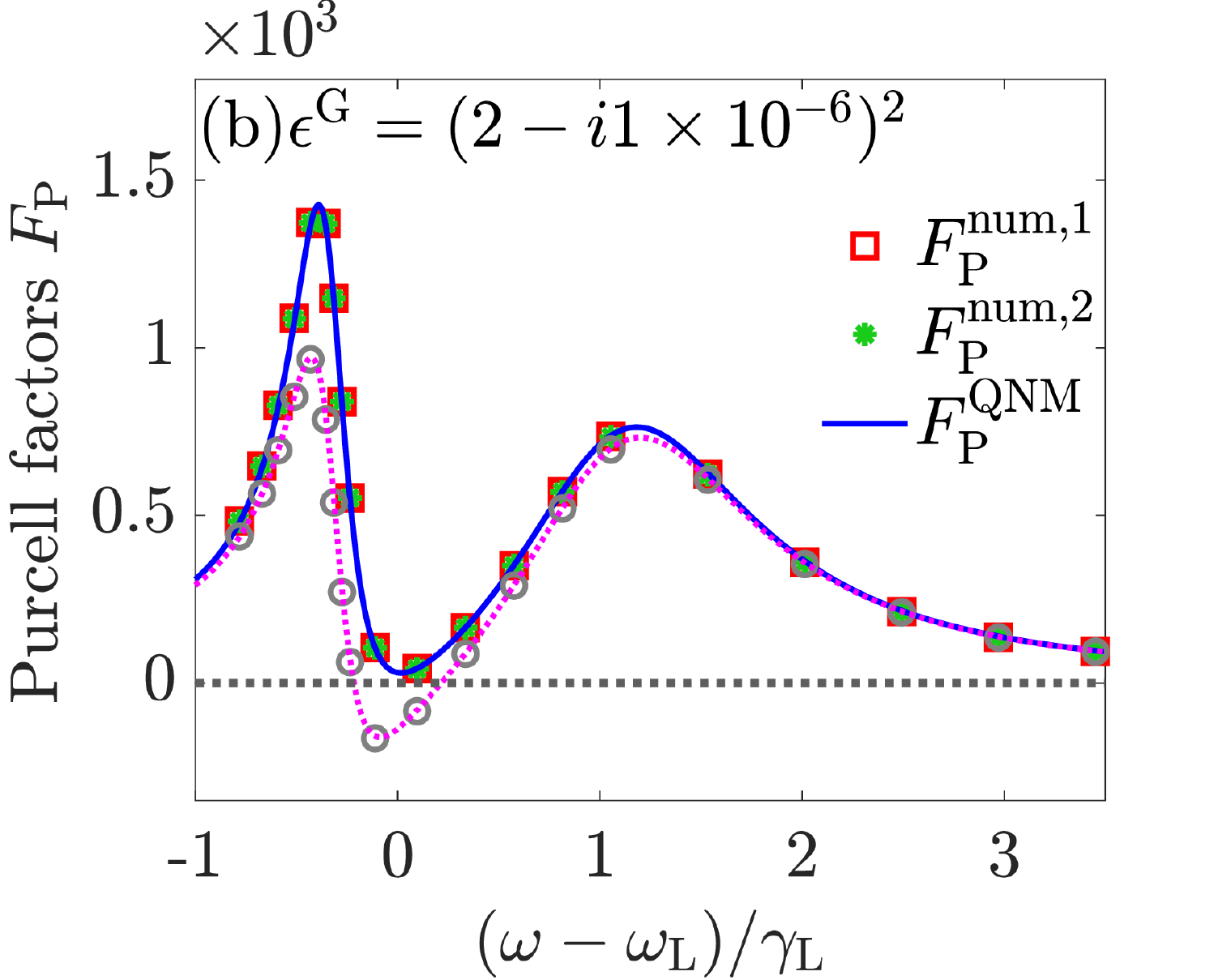}   
    \includegraphics[width = 0.89\columnwidth]{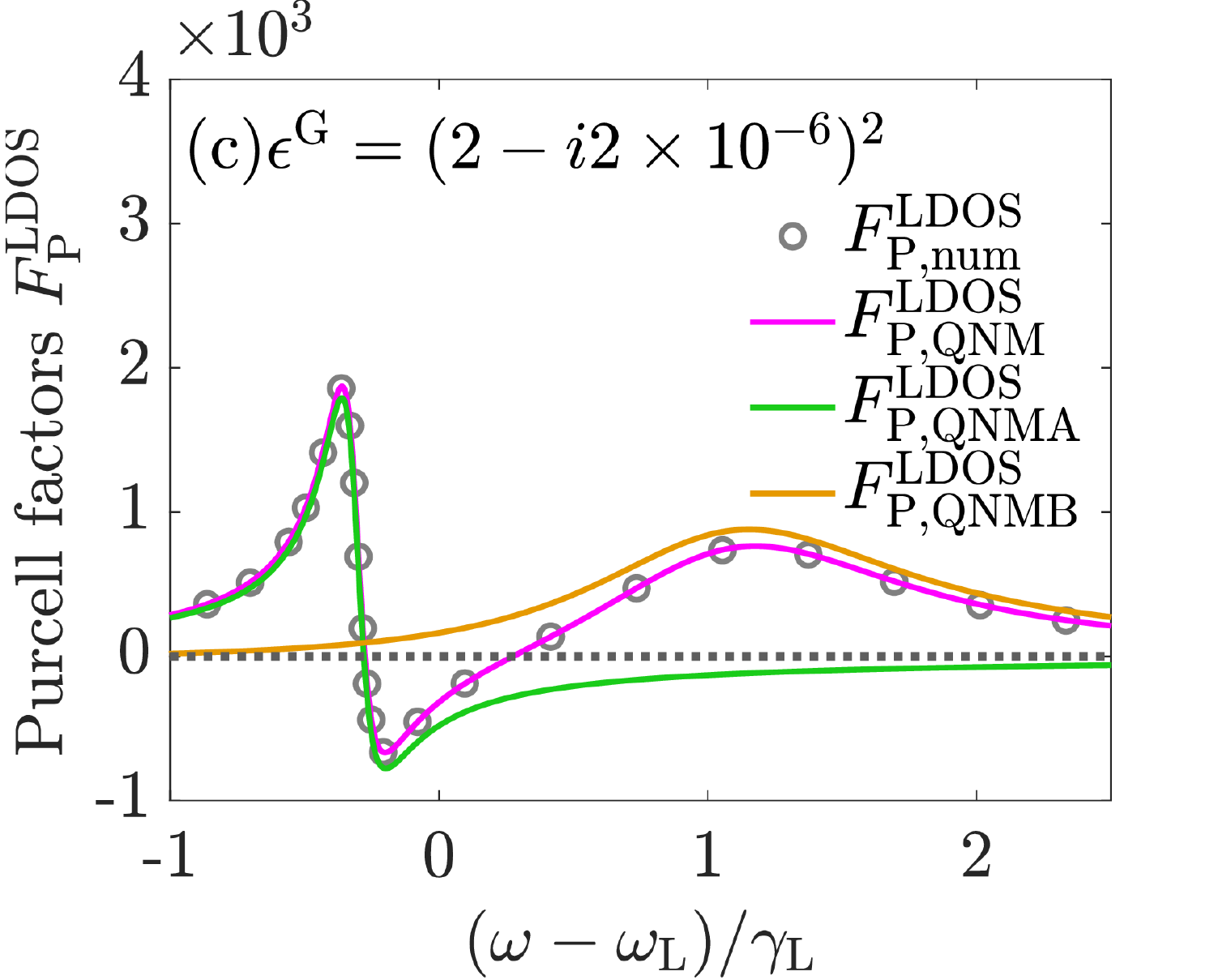}
    \includegraphics[width = 0.89\columnwidth]{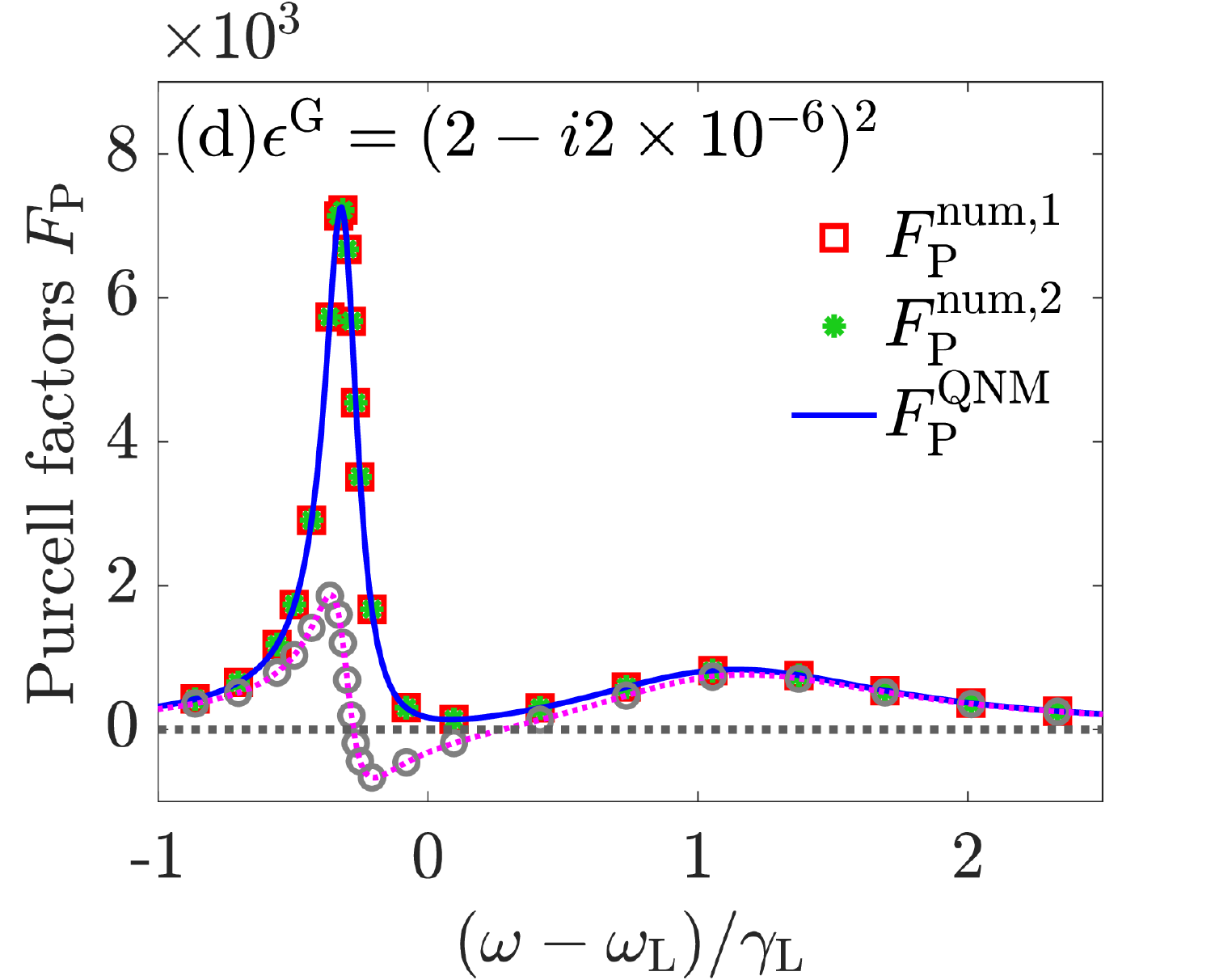}      
    \caption{{(a) LDOS Purcell factor for a dipole placed inside the small dashed circle within lossy disk (Fig.~\ref{fig: sche_loc_lossgain} (a)). $r_{\rm c}=5~$nm and $\epsilon_{\rm c}=1.0$. The gap distance is $d_{\rm gap}=1155~$nm and the permittivity of gain disk is set as $\epsilon^{\rm G}=(2-i1\times10^{-6})^2$. There are two dominant QNMs (see Fig.~\ref{fig: sche_loc_lossgain} (b,c) for mode distributions), whose contributions to the LDOS Purcell factors are described by the green and orange curves. There are very good agreements between the classical QNMs results $F^{\rm LDOS}_{\rm P,QNM}$ [Eq.~\eqref{eq: FP_QNM_LDOS}, magenta curve] and numerical LDOS results $F^{\rm LDOS}_{\rm P,num}$ [Eq.~\eqref{eq: FP_num_LDOS}, grey circles]. The dotted grey horizontal line indicates the value of $0$. However, negative LDOS Purcell factors are found in certain frequency range. (b) Corresponding total net-positive Purcell factors for the case shown in (a), where the corrected numerical Purcell factors $F_{\rm P}^{\rm num,1/2}$ [Eqs.~\eqref{eq: FP_num_mod1} and \eqref{eq: FP_num_mod2}, red squares and green asterisks], agree very well with the Purcell factors $F_{\rm P}^{\rm QNM}$~($=F_{\rm P,QNM}^{\rm class}=F_{\rm P}^{\rm quant}$, solid blue curves, Eq.~\eqref{eq: FP_cQNM_volsurf_2}/Eq.~\eqref{eq: PF_quan}). (c) Similar to (a), but with $\epsilon^{\rm G}=(2-i2\times10^{-6})^2$. (d) Corresponding corrected Purcell factors for the case shown in (c).}
    }
    \label{fig: PF_lossgain_Leps1G}
\end{figure*}

\begin{figure*}[htb]
    \centering
    \includegraphics[width = 0.89\columnwidth]{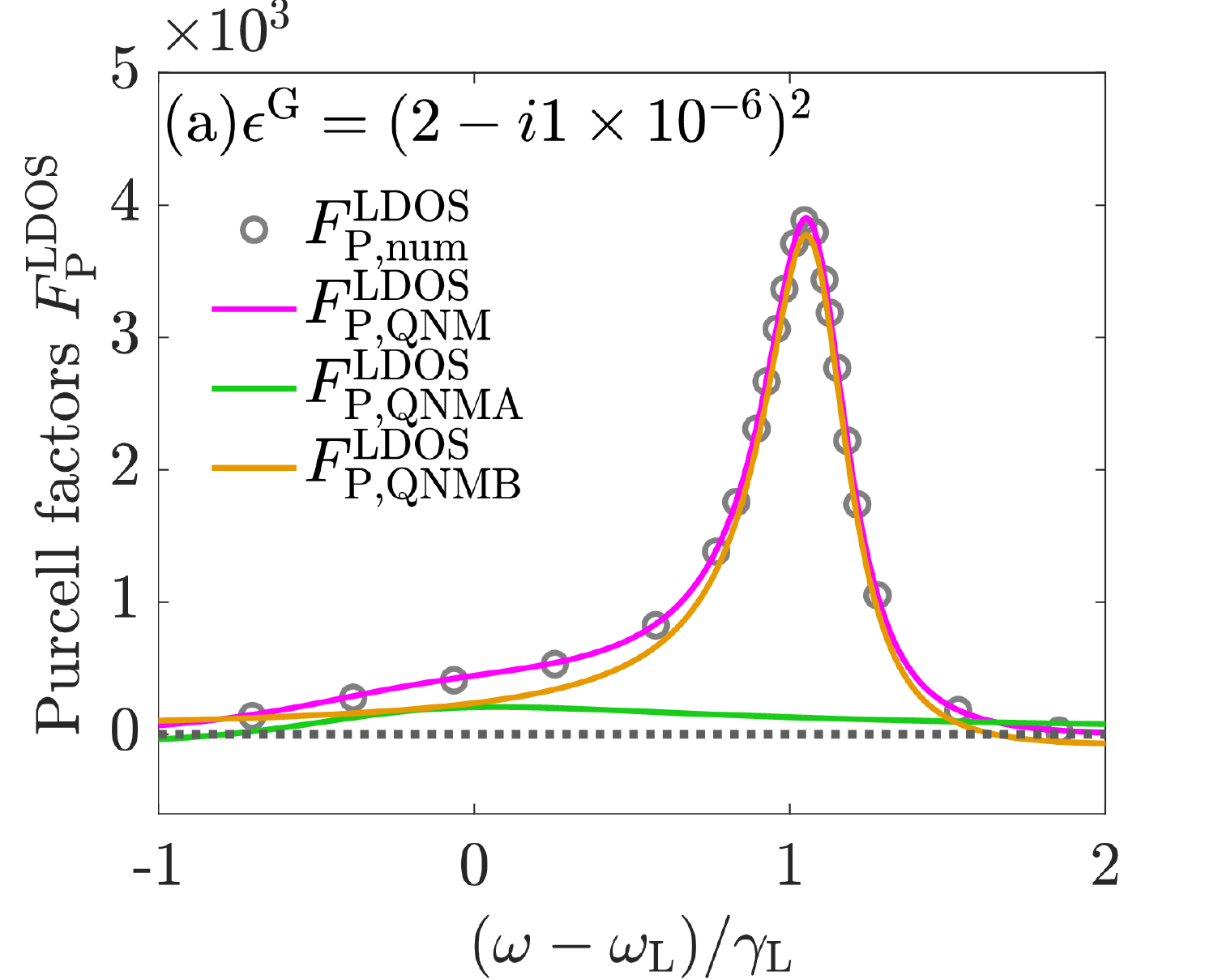}
    \includegraphics[width = 0.89\columnwidth]{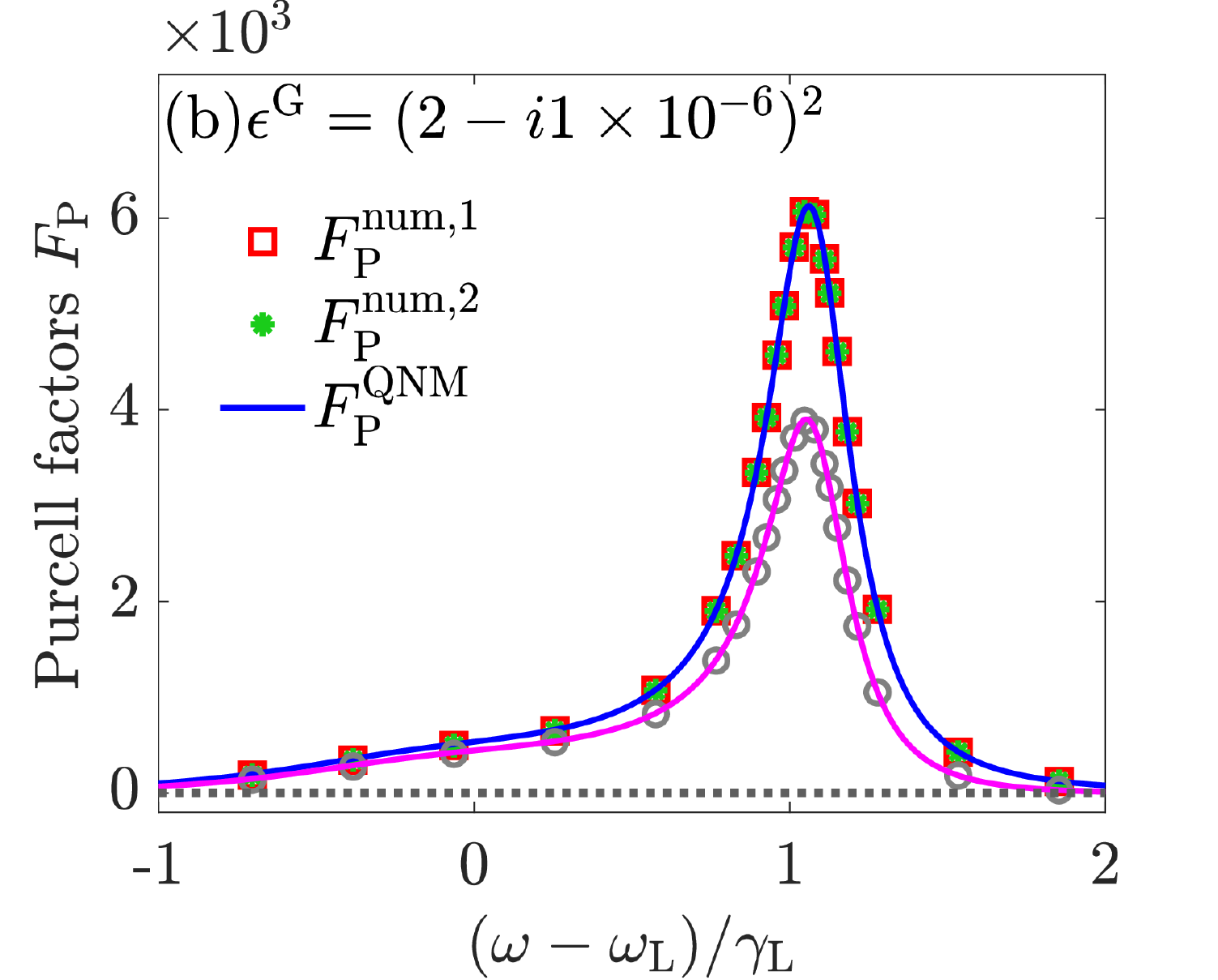}   
    \includegraphics[width = 0.89\columnwidth]{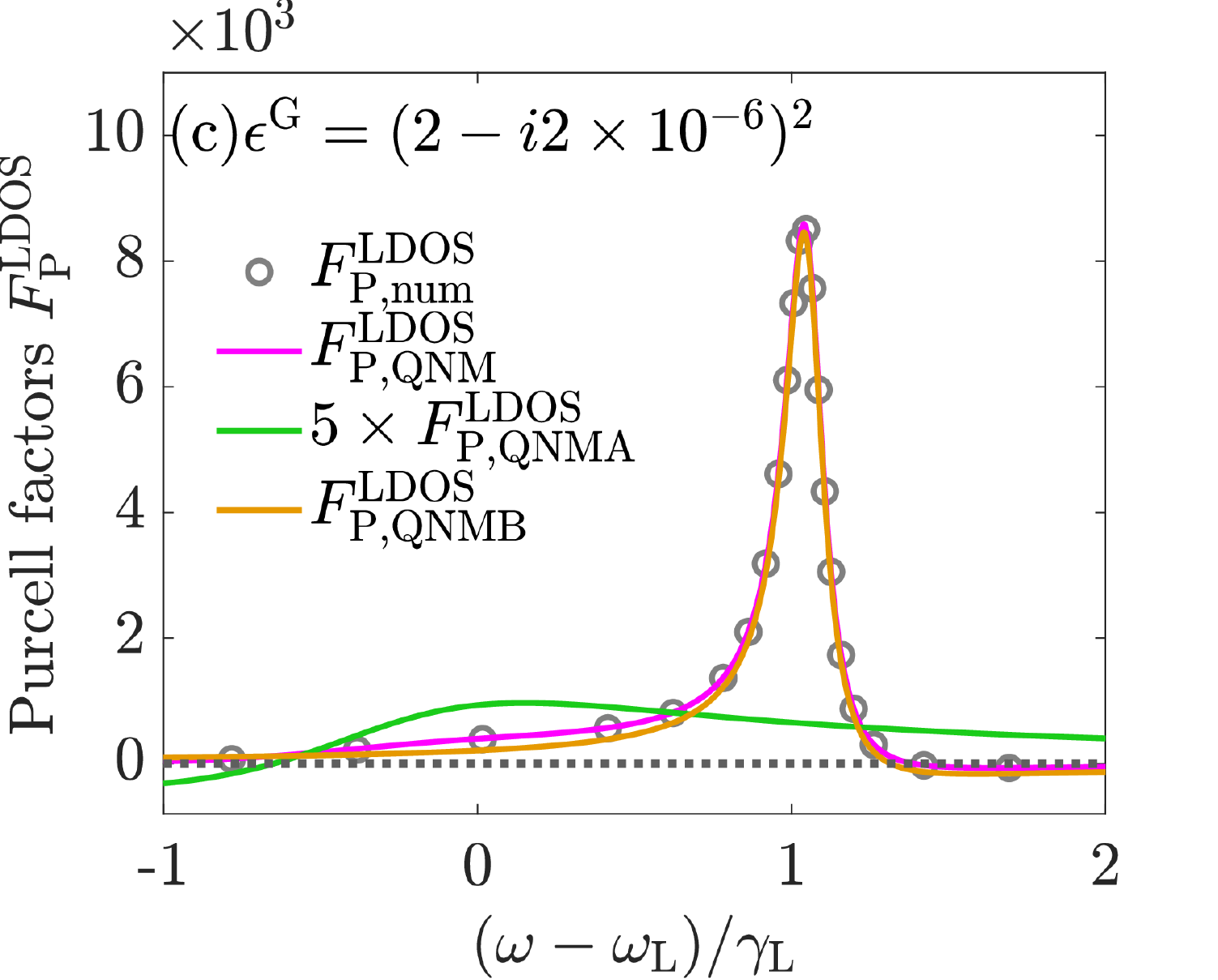}
    \includegraphics[width = 0.89\columnwidth]{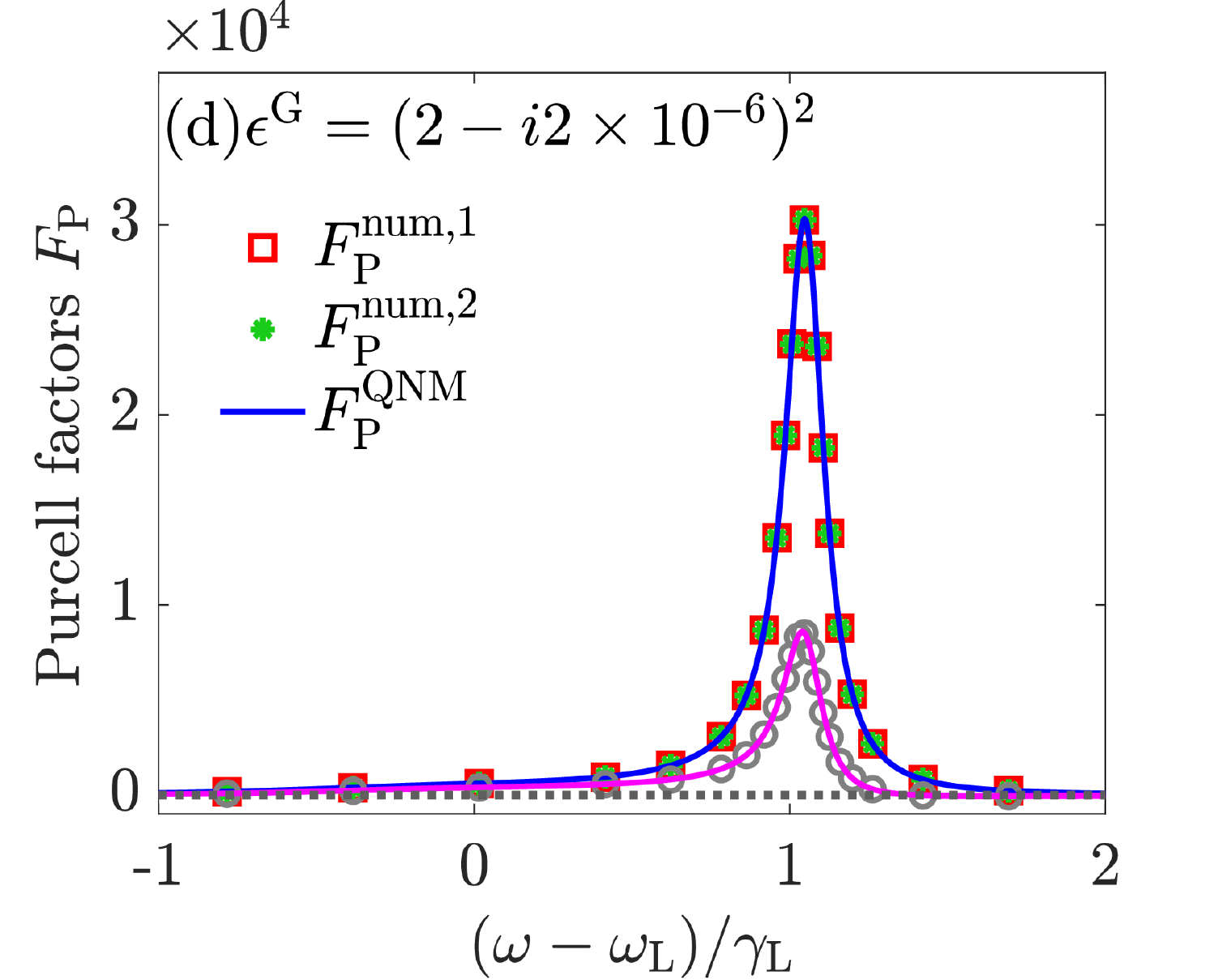}      
    \caption{{(a) Similar to Fig.~\ref{fig: PF_lossgain_Leps1G} (a), but for a dipole placed in the dashed circle within the gain disk (see Fig.~\ref{fig: sche_loc_lossgain} (d)). Note that the LDOS Purcell factors are negative in a certain frequency range (not the range we are showing here). (b) Corresponding corrected net-positive Purcell factors for the case shown in (a). (c) Similar to (a), but with $\epsilon^{\rm G}=(2-i2\times10^{-6})^2$. The results $F_{\rm P,QNMA}^{\rm LDOS}$ from QNM A is multiplied by $5$ for better display.
    (d) Corresponding corrected Purcell factors for the case shown in (c).}
    }
    \label{fig: PF_lossgain_LGeps1}
\end{figure*}

The corresponding Purcell factors for this case, with the real cavity (circle) in the lossy disk and $\epsilon^{\rm G}=(2-i1\times10^{-6})^2$) are shown in Fig.~\ref{fig: PF_lossgain_Leps1G}(a) and (b). 

With local field effects included 
with the calculation of the QNMs,
the LDOS Purcell factors are obtained through
\begin{align}
\mathbf{G}^{\rm QNM}(\mathbf{r}_{0},\mathbf{r}_{0},\omega)\approx & \mathbf{G}^{\rm QNM}_{\rm A}(\mathbf{r}_{0},\mathbf{r}_{0},\omega) + \mathbf{G}^{\rm QNM}_{\rm B}(\mathbf{r}_{0},\mathbf{r}_{0},\omega),\\
\mathbf{G}^{\rm QNM}_{\rm A}(\mathbf{r}_{0},\mathbf{r}_{0},\omega)= &A_{\rm A}(\omega)\tilde{\mathbf{f}}_{\rm A}(\mathbf{r}_{0})\tilde{\mathbf{f}}_{\rm A}(\mathbf{r}_{0}),\\
\mathbf{G}^{\rm QNM}_{\rm B}(\mathbf{r}_{0},\mathbf{r}_{0},\omega)= &A_{\rm B}(\omega)\tilde{\mathbf{f}}_{\rm B}(\mathbf{r}_{0})\tilde{\mathbf{f}}_{\rm B}(\mathbf{r}_{0}),\\
F_{\rm P,QNM}^{\rm LDOS}(\mathbf{r}_0,\omega)=&1+\frac{\mathbf{d}\cdot{\rm Im}[\mathbf{G}^{\rm QNM}(\mathbf{r}_{0},\mathbf{r}_{0},\omega)]\cdot\mathbf{d}}{\mathbf{d}\cdot{\rm Im}[\mathbf{G}_{\rm hom}(\omega)]\cdot\mathbf{d}},\label{eq: FP_QNM_LDOS_loc_lossgain}\\
F_{\rm P,QNMA}^{\rm LDOS}(\mathbf{r}_0,\omega)=&\frac{\mathbf{d}\cdot{\rm Im}[\mathbf{G}^{\rm QNM}_{\rm A}(\mathbf{r}_{0},\mathbf{r}_{0},\omega)]\cdot\mathbf{d}}{\mathbf{d}\cdot{\rm Im}[\mathbf{G}_{\rm hom}(\omega)]\cdot\mathbf{d}},\label{eq: FP_QNMA_LDOS_loc_lossgain}\\
F_{\rm P,QNMB}^{\rm LDOS}(\mathbf{r}_0,\omega)=&\frac{\mathbf{d}\cdot{\rm Im}[\mathbf{G}^{\rm QNM}_{\rm B}(\mathbf{r}_{0},\mathbf{r}_{0},\omega)]\cdot\mathbf{d}}{\mathbf{d}\cdot{\rm Im}[\mathbf{G}_{\rm hom}(\omega)]\cdot\mathbf{d}},\label{eq: FP_QNMB_LDOS_loc_lossgain}
\end{align}
where 
$A_{\rm A/B}(\omega)=\omega/[2(\tilde{\omega}_{\rm A/B}-\omega)]$, and $\tilde{\mathbf{f}}_{\rm A/B}$ and $\tilde{\omega}_{\rm A/B}$ are the two dominant coupled QNMs and the corresponding angular eigenfrequencies.

In Fig.~\ref{fig: PF_lossgain_Leps1G}(a),
we show $F_{\rm P,QNMA/B}^{\rm LDOS}(\mathbf{r}_0,\omega)$ (Eq.~\eqref{eq: FP_QNMA_LDOS_loc_lossgain} and Eq.~\eqref{eq: FP_QNMB_LDOS_loc_lossgain}) from the two coupled QNMs, with the solid green curve and orange curve. The total contribution $F_{\rm P,QNM}^{\rm LDOS}\mathbf{r}_0,\omega)$ [magenta curve, Eq.~\eqref{eq: FP_QNM_LDOS_loc_lossgain}] shows very good agreement with the numerical full-dipole LDOS results $F_{\rm P,num}^{\rm LDOS}$ (grey circles, Eq.~\eqref{eq: FP_num_LDOS}), which verifies that the approximation of two QNMs model worked well in the frequency region that we are considering, even including local field effects. In the presence of gain, similar to before, one can clearly notice that the LDOS Purcell factors are negative in a certain frequency range, which required us to employ the fixes discussed previously (classically and/or quantum mechanically). 

The corresponding {\it corrected Purcell factors} are shown in Fig.~\ref{fig: PF_lossgain_Leps1G}(b). The blue curve represent Purcell factors $F_{\rm P}^{\rm QNM}=F_{\rm P,QNM}^{\rm class}=F_{\rm P}^{\rm quant}$, which are now net positive (the LDOS results are also shown for comparison, see dotted magenta curve and grey circles). Note, as mentioned before, the form of $F_{\rm P,QNM}^{\rm class}$ shown in Eq.~\eqref{eq: FP_cQNM_volsurf_2} is identical 
to $F_{\rm P}^{\rm quant}$ (Eq.~\eqref{eq: PF_quan}) (we are showing this as the solid blue curve). The approximation of the second form of $F_{\rm P,QNM}^{\rm class}$ shown in Eq.~\eqref{eq: FP_cQNM_volsurf_1} also worked for our examples (not shown, but equal to the above form) as the nonradiative contributions dominate. Moreover, based on power flow computations, the full dipole results $F_{\rm P}^{\rm num,1/2}$ (Eqs.~\eqref{eq: FP_num_mod1} and \eqref{eq: FP_num_mod2}, red squares and green asterisks) agreed very well with $F_{\rm P}^{\rm QNM}$, which verified that our general conclusions ($\rm LDOS+gain=nloss+rloss$) also works in the case where the emitter is placed within the lossy resonators.

To show this more generality, we also investigated some additional cases. Working again with the circular region in the lossy disk, the permittivity of the gain disk is now changed to $\epsilon_{\rm G}=(2-i2\times10^{-6})^2$. The LDOS Purcell factors are shown in Fig.~\ref{fig: PF_lossgain_Leps1G}(c), and the corrected Purcell factors are shown in Fig.~\ref{fig: PF_lossgain_Leps1G}(d). Once again, the underestimated LDOS Purcell factors can be negative in a certain frequency range, but the corrected Purcell factors are net positive. The fixed classical results based on power flow or QNMs not only matched very well with the quantum mechanical results, but also provide us with an alternative way to picture the process
of spontaneous emission, namely, the sum of contributions from nonradiative loss and radiative loss will give us exactly the same answer (rate) as the LDOS contribution plus the addition gain contribution. Moreover, we find that this conclusion still holds when the dipole is placed within the gain disk [Fig.~\ref{fig: sche_loc_lossgain}(d)] as shown in Fig.~\ref{fig: PF_lossgain_LGeps1}, where the cases with $\epsilon_{\rm G}=(2-i1\times10^{-6})^2$ [see Figs.~\ref{fig: PF_lossgain_LGeps1}(a) and ~\ref{fig: PF_lossgain_LGeps1}(b)] and $\epsilon_{\rm G}=(2-i2\times10^{-6})^2$ [see Figs.~\ref{fig: PF_lossgain_LGeps1}(c) and ~\ref{fig: PF_lossgain_LGeps1}(d)] are studied.

}

\section{Discussion and connection to quantized quasinormal mode results in the bad cavity limit}

We have shown that the contributions to the SE rate can be obtained from an LDOS term plus a nonlocal gain term, or alternatively from a nonlocal loss term plus the radiative decay to the far field, i.e.,  $\Gamma_{\rm class}^{\rm SE}=\Gamma^{\rm LDOS}+\Gamma_{\rm class}^{\rm gain}$ or $\Gamma_{\rm class}^{\rm SE}=\Gamma_{\rm class}^{\rm rloss}+\Gamma_{\rm class}^{\rm nloss}$ from 
Eq.~\eqref{eq: FP_mod2_decay} and Eq.~\eqref{eq: FP_mod1_decay}.
Next, we will connect these classical results with the ones from rigorous quantized QNM theory in the bad cavity limit.

Note in the coupled resonator example discussed above, the contribution to the far field decay is negligible, and
thus, 
Eq.~\eqref{eq: FP_mod1_decay} could be approximated as $\Gamma_{\rm class}^{\rm SE}\approx\Gamma_{\rm class}^{\rm nloss}$, as verified above.
However, to be general in our theory, below we will formulate the quantum theory with the general radiative (far field) and non-radiative (within lossy region) contributions.

In a  quantized QNM picture, one 
starts by computing the quantum $S$ parameters (defined below), which enter the relevant quantum master equations.
The matrix $S_{\mu\eta}$ is a semi-positive definite Hermitian
overlap matrix between different QNMs,
and is not a {Kronecker} delta 
as in the case of simple normal modes,
e.g., for a closed cavity.
These factors are necessary to construct a meaningful Fock space with modal losses and gain (QNMs)~\cite{franke_quantization_2019,franke_quantized_2020,PhysRevA.105.023702}.
Using the QNM master equation, a quantum-classical correspondence can be derived
by taking a bad
cavity limit.

The QNM master equation was 
 originally derived for lossy media only~\cite{franke_quantization_2019}.
 Later, 
in Ref.~\cite{PhysRevA.105.023702},
two forms of quantization were presented when including gain: (i) using separated operators for loss (which includes both radiative and nonradiative contributions in general) and gain, or (ii) using combined QNM operators.
Since both approaches yield the same bad cavity limit rates, below we will focus on the first approach. {Furthermore, below,  for ease of notation, we will drop the operator hat on the QNM and emitter operators (except for the electric field operator) and assume their operator character as implicit.}

Using  separated operators for loss and gain, there are two QNM contributions to the  electric field operator,
\begin{align}
\hat{\mathbf{E}}(\mathbf{r})=\hat{\mathbf{E}}^{\rm L}_{\rm QNM}(\mathbf{r})+\hat{\mathbf{E}}^{\rm G}_{\rm QNM}(\mathbf{r}), 
\end{align}
where the lossy/gain (L/G) parts are 
\begin{align}
\hat{\mathbf{E}}^{\mathrm{L}}_{\rm QNM}(\mathbf{r})&=i\sum_\mu\sqrt{\frac{\hbar\omega_\mu}{2\epsilon_0}}\tilde{\mathbf{f}}_\mu^{\mathrm{s,L}}(\mathbf{r})a_{{\rm L}\mu}+\mathrm{H.a.}\label{eq: ESymLoss}, \\
\hat{\mathbf{E}}^{\mathrm{G}}_{\rm QNM}(\mathbf{r})&=i\sum_\mu\sqrt{\frac{\hbar\omega_\mu}{2\epsilon_0}}\tilde{\mathbf{f}}_\mu^{\mathrm{s,G}}(\mathbf{r})a_{{\rm G}\mu}^\dagger+\mathrm{H.a.}\label{eq: ESymGain}.,
\end{align}
and $\mathrm{H.a.}$ represents Hermitian adjoint.
The constructed annihilation and creation operators for both {\it loss} Fock {space} ($a_{{\rm L}\mu}$ and $a_{{\rm L}\mu}^{\dagger}$) and {\it gain} Fock {space} ($a_{{\rm G}\mu}$ and $a_{{\rm G}\mu}^{\dagger}$), are closely related to the {loss- and gain}-assisted QNM operators $\tilde{a}_{{\rm L}}$ and $\tilde{a}_{{\rm G}}$ through a symmetrization transformation, 
\begin{align}
a_{{\rm L}\mu} &= \sum_\eta \left[\left(\mathbf{S}^\mathrm{L}\right)^{1/2}\right]_{\mu\eta}\tilde{a}_{{\rm L}\eta},\\
a_{{\rm G}\mu} &= \sum_\eta \left[\left(\mathbf{S}^\mathrm{G}\right)^{1/2}\right]_{\mu\eta}\tilde{a}_{{\rm G}\eta},
\end{align}
where QNM operators $\tilde{a}_{{\rm L(G)}}$ satisfy %
\begin{align}
[\tilde{a}_{{\rm L}\mu},\tilde{a}^{\dagger}_{{\rm L}\eta}]&\equiv S_{\mu\eta}^{\rm L}=S^{\mathrm{rloss}}_{\mu\eta}+S^{\mathrm{nloss}}_{\mu\eta},\\
[\tilde{a}_{{\rm G}\mu},\tilde{a}^{\dagger}_{{\rm G}\eta}]&\equiv S_{\mu\eta}^{\rm G}.
\end{align}

The required quantum $S$ parameters are defined from:
\begin{align}
S^{\mathrm{rloss}}_{\mu\eta} &=\int_0^\infty{\mathrm d}\omega\frac{2A_\mu(\omega)A_\eta^*(\omega)}{\pi\sqrt{\omega_\mu\omega_\eta}}[I^{\rm rloss}_{\mu\eta}(\omega)+I^{\rm rloss*}_{\eta\mu}(\omega)] \label{eq:SLradGeneral},\\
S_{\mu\eta}^\mathrm{nloss}&=\int_0^\infty{\mathrm d}\omega\frac{2A_\mu(\omega)A_\eta^*(\omega)}{\pi\sqrt{\omega_\mu\omega_\eta}}I^{\rm nloss}_{\mu\eta}(\omega),\label{eq:SLnradGeneral}\\
S_{\mu\eta}^\mathrm{G}&=\int_0^\infty{\mathrm d}\omega\frac{2A_\mu^*(\omega)A_\eta(\omega)}{\pi\sqrt{\omega_\mu\omega_\eta}}I^{\rm G}_{\mu\eta}(\omega),\label{eq:SGGeneral}
\end{align}
with 
\begin{align}
I^{\rm rloss}_{\mu\eta}(\omega)&=\frac{1}{2\epsilon_0\omega}\oint_{\mathcal{S}} {\mathrm d}\mathbf{r}\left[\tilde{\mathbf{H}}_{\mu}(\mathbf{r},\omega)\times\hat{\mathbf{n}}\right]\cdot\tilde{\mathbf{F}}_{\eta}^*(\mathbf{r},\omega),\\
I^{\rm nloss}_{\mu\eta}(\omega)&=\int_{V_{\rm L}} {\mathrm d}\mathbf{r}~\epsilon_{\rm Im}^{\rm L}(\mathbf{r},\omega)\tilde{\mathbf{f}}_\mu(\mathbf{r})\cdot\tilde{\mathbf{f}}_\eta^*(\mathbf{r}),\\
I^{\rm G}_{\mu\eta}(\omega)&=\int_{V_{\rm G}} {\mathrm d}\mathbf{r}\left|\epsilon_{\rm Im}^{\rm G}(\mathbf{r},\omega)\right|\tilde{\mathbf{f}}_\mu^*(\mathbf{r})\cdot\tilde{\mathbf{f}}_\eta(\mathbf{r}),
\end{align}
where $\tilde{\mathbf{F}}_{\mu}(\mathbf{r},\omega)$ ($\tilde{\mathbf{H}}_{\mu}(\mathbf{r},\omega)=\frac{1}{i\omega\mu_{0}}\nabla\times\tilde{\mathbf{F}}_{\mu}(\mathbf{r},\omega)$) is the regularized electric (magnetic) QNM~\cite{ge_design_2014,ren_near-field_2020}, and $\hat{\mathbf{n}}$ denotes the unit vector normal to surface $\mathcal{S}$ (a far field closed surface), pointing outward.
The term $V_{\rm L(G)}$ represents the region with material loss (gain); $S_{\mu\eta}^{\rm rloss}$ represents the radiative loss part to the far field region; and $S_{\mu\eta}^{\rm nloss}$ represents the nonraditive absorption within lossy region $V_{\rm L}$; finally, $S_{\mu\eta}^{\rm G}$ represents  the amplification contribution within the gain region $V_{\rm G}$.

The symmetrized QNM functions in Eq.~\eqref{eq: ESymLoss} and Eq.~\eqref{eq: ESymGain} are defined as
\begin{align}
\tilde{\mathbf{f}}_\mu^{\mathrm{s,L}}(\mathbf{r})&=\sum_\eta \left[\left(\mathbf{S}^\mathrm{L}\right)^{1/2}\right]_{\eta\mu}\tilde{\mathbf{f}}_\eta(\mathbf{r})\sqrt{\frac{\omega_\eta}{\omega_\mu}},\\
\tilde{\mathbf{f}}_\mu^{\mathrm{s,G}}(\mathbf{r})&=\sum_\eta \left[\left(\mathbf{S}^\mathrm{G}\right)^{1/2}\right]_{\mu\eta}\tilde{\mathbf{f}}_\eta(\mathbf{r}) \sqrt{\frac{\omega_\eta}{\omega_\mu}}.
\end{align}

{The full Lindblad QNM master equation can be written as 
\begin{equation}
    \partial_t \rho = -\frac{i}{\hbar}[H_{\rm em} + H_{\rm a} + H_I,\rho_{\rm a}]+\mathcal{L}_{\rm em}\rho\label{eq: masterQNM}
\end{equation}
where $H_{\rm a}=\hbar\omega_{\rm a}\sigma^{\dagger}\sigma^{-}$  (raising and lowering operators $\sigma^{\pm}$) is the energy of the two-level system and $H_I$ is the dipole-field interaction Hamiltonian using the loss and gain QNM fields, defined in Eq.~\eqref{eq: ESymLoss} and Eq.~\eqref{eq: ESymGain}, respectively. Furthermore, $H_{\rm em}$ is the QNM photon energy, and $\mathcal{L}_{\rm em}$ is the QNM Lindblad dissipator. For more details on the derivation of the QNM master equation in the presence of gain and loss, we refer to Ref.~\onlinecite{PhysRevA.105.023702}. From this point on, we concentrate on} the weak coupling limit, {where the QNM decay rates are much larger compared to the dipole-field coupling energy. Applying the bad cavity limit of Eq.~\eqref{eq: masterQNM}, we arrive at} the TLS master equation for the atomic density operator $\rho_{\rm a}={\rm tr}_{\rm em}\rho$ within the quantized QNM models is obtained as~\cite{PhysRevA.105.023702}:
\begin{align}
    \partial_t\rho_{\rm a}&=-\frac{i}{\hbar}[H_{\rm a},\rho_{\rm a}]+ \frac{\Gamma^{\rm B}}{2}\mathcal{D}[\sigma^-]\rho_{\rm a}\nonumber\\
    &+\frac{\Gamma^{\rm SE}_{\rm bad}}{2}\mathcal{D}[\sigma^-]\rho_{\rm a}+\frac{\Gamma^{\rm gain}_{\rm bad}}{2}\mathcal{D}[\sigma^+]\rho_{\rm a},\label{eq: BadCavMaster}
\end{align}
 and the Lindblad dissipator,
\begin{align}
   \mathcal{D}[A]\rho_{\rm a}=2A\rho_{\rm a}A^{\dagger}- \rho_{\rm a}A^{\dagger}A-A^{\dagger}A\rho_{\rm a}.
\end{align}
The medium-dependent SE rate is $\Gamma^{\rm SE}_{\rm bad}=\Gamma^{\rm rloss}_{\rm bad}+\Gamma^{\rm nloss}_{\rm bad}$, similar to the the classical separation shown in Eq.~\eqref{eq: FP_mod1_decay}.
In the quantum derivation, the radiative and nonradiative contributions are
\begin{align}
    \Gamma^{\rm rloss}_{\rm bad}(\mathbf{r}_0,{\omega_{\rm a}})=\sum_{\mu,\eta}\tilde{g}_\mu S_{\mu\eta}^{\rm rloss}\tilde{g}_{\eta}^*\frac{i(\omega_\mu-\omega_{\eta})+(\gamma_{\mu }+\gamma_{\eta})}{(\Delta_{\mu a}-i\gamma_{\mu })(\Delta_{\eta a}+i\gamma_{\eta})}\label{eq: Gamma_far_bad},\\
    \Gamma^{\rm nloss}_{\rm bad}(\mathbf{r}_0,{\omega_{\rm a}})=\sum_{\mu,\eta}\tilde{g}_\mu S_{\mu\eta}^{\rm nloss}\tilde{g}_{\eta}^*\frac{i(\omega_\mu-\omega_{\eta})+(\gamma_{\mu }+\gamma_{\eta})}{(\Delta_{\mu a}-i\gamma_{\mu })(\Delta_{\eta a}+i\gamma_{\eta})}\label{eq: Gamma_loss_bad},
\end{align}
where $S_{\mu\eta}^{\rm rloss}$ and $S_{\mu\eta}^{\rm nloss}$ are defined from Eq.~\eqref{eq:SLradGeneral} and Eq.~\eqref{eq:SLnradGeneral}, 
 $\Delta_{\mu{\rm a}/\eta{\rm a}}=\omega_{\mu/\eta}-\omega_{\rm a}$ gives the frequency detuning between the QNM and the emitter, and the emitter-QNM coupling strength is given by $\tilde{g}_{\mu/\eta}=\sqrt{\frac{\omega_{\mu/\eta}}{2\varepsilon_0\hbar}}\mathbf{d}\cdot\tilde{\mathbf{f}}_{\mu/\eta}(\mathbf{r}_{0})$.

For the specific resonator example considered above, the nonradiative part dominates, so we only have to consider $S^{\rm nloss}_{\mu\eta}$ and $\Gamma^{\rm nloss}_{\rm bad}$, as the far field radiative contribution can be safely neglected  (i.e.,  $S^{\rm rloss}_{\mu\eta}$ and $\Gamma^{\rm rloss}_{\rm bad}$ are negligible). Then, in the bad cavity limit, the quantum result for our resonator example can be approximated as $\Gamma^{\rm SE}_{\rm bad}(\mathbf{r}_{0},\omega_{\rm a})\approx\Gamma^{\rm nloss}_{\rm bad}(\mathbf{r}_{0},\omega_{\rm a})$; this can be compared with the same classical approximation $\Gamma_{\rm class}^{\rm SE}(\mathbf{r}_0,\omega)\approx\Gamma_{\rm class}^{\rm nloss}(\mathbf{r}_0,\omega)$.
Note again here the quantum result is at the frequency $\omega_{\rm a}$ of the emitter, and the classical result is at the linear frequency $\omega$ of interest, but it is clear $\omega=\omega_{\rm a}$ when comparing the two.

Moreover, the gain-induced pump rate in Eq.~\eqref{eq: BadCavMaster} is given as
\begin{equation}
    \Gamma^{\rm gain}_{\rm bad}(\mathbf{r}_0,\omega_{\rm a})=\sum_{\mu,\eta}\tilde{g}_\mu S_{\eta\mu }^{\rm G}\tilde{g}_{\eta}^*\frac{i(\omega_\mu-\omega_{\eta})+(\gamma_{\mu }+\gamma_{\eta})}{(\Delta_{\mu a}-i\gamma_{\mu })(\Delta_{\eta a}+i\gamma_{\eta})}\label{eq: GammaPump},
\end{equation}
where the $S$ parameters are shown in Eq.~\eqref{eq:SGGeneral}.
Note that,  as shown in Ref.~\onlinecite{PhysRevA.105.023702}, the difference between $\Gamma^{\rm SE}_{\rm bad}(\mathbf{r}_0,\omega_{\rm a})$ and $\Gamma^{\rm gain}_{\rm bad}(\mathbf{r}_0,\omega_{\rm a})$ is directly related to the projected LDOS SE rate, through: 
\begin{equation}
\Gamma^{\rm SE}_{\rm bad}(\mathbf{r}_0,\omega_{\rm a})-\Gamma^{\rm gain}_{\rm bad}(\mathbf{r}_0,\omega_{\rm a})=\Gamma^{\rm LDOS}(\mathbf{r}_0,\omega_{\rm a}),
\end{equation}
with an analogous separation also shown in the classical results, from  Eq.~\eqref{eq: FP_mod2_decay}.

Next, in order to show a clearer quantum-classical correspondence, we wish to connect the quantum result, $\Gamma_{\rm bad}^{\rm gain}$ shown in Eq.~\eqref{eq: GammaPump}, with
the classical result, $\Gamma_{\rm class}^{\rm gain}$ described in Eq.~\eqref{eq: Gamma_Gvol}. By
substituting $S_{\mu\eta}^{\rm G}$ [Eq.~\eqref{eq:SGGeneral}] into $\Gamma_{\rm bad}^{\rm gain}$ [Eq.~\eqref{eq: GammaPump}], we have 
\begin{widetext}
\begin{align}\label{eq: gamma_bad_gain_con1}
    \Gamma^{\rm gain}_{\rm bad}(\mathbf{r}_0,\omega_{\rm a})=\frac{1}{2\epsilon_0\hbar}\sum_{\mu,\eta}\mathbf{d}\cdot\left[\int_{V_{\rm G}} {\mathrm d}{\bf r}\int_0^\infty{\mathrm d}\omega\frac{2A_\mu(\omega)A_{\eta}^*(\omega)}{\pi}|\epsilon_{\rm Im}^{\rm G}(\mathbf{r},\omega)|\tilde{\mathbf{f}}_\mu(\mathbf{r}_0)\tilde{\mathbf{f}}_\mu(\mathbf{r})\cdot\tilde{\mathbf{f}}_{\eta}^*(\mathbf{r})\tilde{\mathbf{f}}_{\eta}^*(\mathbf{r}_0)\right]\cdot\mathbf{d}\frac{i(\omega_\mu-\omega_{\eta})+(\gamma_{\mu }+\gamma_{\eta})}{(\Delta_{\mu a}-i\gamma_{\mu })(\Delta_{\eta a}+i\gamma_{\eta})}.
\end{align}
Using the definition of $A_\mu(\omega)$, i.e., $A_\mu(\omega)=\omega/(2(\tilde{\omega}_\mu-\omega))$, then we obtain
\begin{equation}\label{eq: gamma_bad_gain_con2}
    \Gamma_{\rm bad}^{\rm gain}(\mathbf{r}_0,\omega_{\rm a})=\frac{2}{\epsilon_0\hbar}\sum_{\mu,\eta}\mathbf{d}\cdot\left[\int_{V_{\rm G}} {\mathrm d}{\bf r}\frac{i}{2\pi}\int_0^\infty{\mathrm d}\omega\frac{\omega^2}{\omega_{a}^2}\frac{\tilde{\omega}_\mu-\tilde{\omega}_{\eta}^*}{(\tilde{\omega}_\mu-\omega)(\tilde{\omega}_{\eta}^*-\omega)}|\epsilon_{\rm Im}^{\rm G}(\mathbf{r},\omega)|A_\mu(\omega_a)A_{\eta}^*(\omega_a)\tilde{\mathbf{f}}_\mu(\mathbf{r}_0)\tilde{\mathbf{f}}_\mu(\mathbf{r})\cdot\tilde{\mathbf{f}}_{\eta}^*(\mathbf{r})\tilde{\mathbf{f}}_{\eta}^*(\mathbf{r}_0)\right]\cdot\mathbf{d},
\end{equation}
which can  be written as [$\mathbf{G}_\mu(\mathbf{r}_0,\mathbf{r},\omega_a)=A_{\mu}(\omega_{\rm a})\tilde{\mathbf{f}}_{\mu}(\mathbf{r}_{0})\tilde{\mathbf{f}}_{\mu}(\mathbf{r})$, $\mathbf{G}(\mathbf{r}_0,\mathbf{r},\omega_a)=\sum_{\mu}\mathbf{G}_\mu(\mathbf{r}_0,\mathbf{r},\omega_a)$]
\begin{equation}
    \Gamma^{\rm gain}_{\rm bad}(\mathbf{r}_0,\omega_{\rm a})=\frac{2}{\hbar\epsilon_0}\sum_{\mu,\eta}\mathbf{d}\cdot\left[\int_{V_{\rm G}} {\mathrm d}\mathbf{r}~K_{\mu\eta}(\mathbf{r})|\epsilon_{\rm Im}^{\rm G}(\mathbf{r},\omega_a)|\mathbf{G}_\mu(\mathbf{r}_0,\mathbf{r},\omega_a)\cdot\mathbf{G}_{\eta}^*(\mathbf{r},\mathbf{r}_0,\omega_a)\right]\cdot\mathbf{d},
\end{equation}
with
\begin{equation}
    K_{\mu\eta}=\frac{i}{2\pi}\int_0^\infty{\mathrm d}\omega\frac{\omega^2|\epsilon_{\rm Im}^{\rm G}(\mathbf
    r,\omega)|}{\omega_{a}^2|\epsilon_{\rm Im}^{\rm G}(\mathbf{r},\omega_a)|}\frac{\tilde{\omega}_\mu-\tilde{\omega}_{\eta}^*}{(\tilde{\omega}_\mu-\omega)(\tilde{\omega}_{\eta}^*-\omega)}.
\end{equation}

Within a pole approximation, we can extend the integral boundaries to $(-\infty,\infty)$  and set $\omega^2|\epsilon_{\rm Im}^{\rm G}(\mathbf{r},\omega)|\approx \omega_a^2|\epsilon_{\rm Im}^{\rm G}(\mathbf{r},\omega_{\rm a})|$,
so that
$K_{\mu\eta} = 1$. 
Finally, this  leads to
\begin{equation}
    \Gamma^{\rm gain}_{\rm bad}(\mathbf{r}_0,\omega_{\rm a})=\frac{2}{\hbar\epsilon_0}\mathbf{d}\cdot\left[\int_{V_{\rm G}} {\mathrm d}{\bf r}|\epsilon_{\rm Im}^{\rm G}(\mathbf{r},\omega_a)|\mathbf{G}(\mathbf{r}_0,\mathbf{r},\omega_a)\cdot\mathbf{G}^*(\mathbf{r},\mathbf{r}_0,\omega_a)\right]\cdot\mathbf{d},
\end{equation}
where we sum over the QNM Green function expansions to get the total Green functions. This is precisely the result we obtain from the classical derivations [$\Gamma_{\rm class}^{\rm gain}$, Eq.~\eqref{eq: Gamma_Gvol}]. A similar connection can be directly made between $\Gamma_{\rm bad}^{\rm rloss}$ [Eq.~\eqref{eq: Gamma_loss_bad}] and $\Gamma_{\rm class}^{\rm rloss}$ [Eq.~\eqref{eq: Gamma_Lvol}]. Moreover, the same arguments can also be made using a pole approximation for the quantum $S$ parameters, as used in Refs.~\cite{ren_near-field_2020,PhysRevA.105.023702}. Thus, we have shown how a quantized QNM approach is completely consistent with 
our classical theory of SE decay in a general loss-gain medium. 
\end{widetext}

\section{Conclusions}
We have presented a corrected form for the classical SE rate and the classical Purcell factor for a dipole emitter in a medium containing a linear amplifier. We have shown how this form recovers a recently presented quantum mechanical form, argued from the viewpoint of specific field operator terms in a Fermi's golden rule.
This classical corrected form compliments the traditional LDOS formula with a nonlocal gain correction.
This work yields a fundamental correction to the meaning of ``radiation reaction'' and extends it to account for additional reaction terms from the gain amplifying part of the medium, which are necessarily nonlocal.

We have also presented an alternative form for the total SE rate with gain media, which is shown to yield equivalent results, in terms of the total material non-radiative loss and the far-field radiative loss, which is valid with and without gain. In such a picture, there is no need to invoke an ill-defined LDOS contribution which may also be negative in such gain media. This work complements the formal quantum theory by offering simpler forms that can easily be checked in a classical Maxwell solver, also yielding 
classical-quantum correspondence. Specific examples were shown for coupled loss-gain resonators at various dipole locations and with different gap separations. Excellent agreement was shown between the various analytical and numerical decay forms, which were also supported by a QNM analysis for the Green function expansions. 

{We have also shown how our general approach can 
model a practical real cavity model, which describes
finite-size dipole emitters inside the loss or gain materials, while also including local field effects. By computing the QNMs in the presence of the real cavities, we have shown an excellent agreement with full dipole scattering simulations, with just one or two QNMs, and also shown how to {\it fix} the LDOS SE rates and Purcell factors to account for gain modifications.}

Finally, we also showed how a fully quantized QNM theory, which
was recently introduced for gain media~\cite{PhysRevA.105.023702}, can be used to make a formal connection back to our modified classical results, in a bad cavity limit. Direct quantum-classical correspondence was confirmed.
{Outside the bad cavity limits, then one can adopt the full quantized QNM
theory using some of the same classical QNM
parameters that we use here. Obviously in such a regime, there is no longer a classical correspondence, and one can explore uniquely quantum optical interactions (such as multi-photon correlation functions). However, clearly one must first recover a classical correspondence in the bad cavity limit, to have confidence that the quantum theory beyond such a limit is accurate and appropriate. In the presence
of linear gain, this is precisely the main goal and accomplishment of this paper. }

\acknowledgements
This work was supported by the Natural Sciences and Engineering Research Council of Canada (NSERC), 
the Canadian Foundation for Innovation (CFI), Queen's University, and the
Alexander von Humboldt Foundation through a Humboldt Research Award.
We also thank 
CMC Microsystems for the provision of COMSOL Multiphysics.

\vspace{0.2cm}

\bibliography{references}

\end{document}